\documentclass[preprint,prd,aps,showpacs,showkeys,nofootinbib]{revtex4}
\usepackage{graphicx}
\usepackage{dcolumn}
\usepackage{bm}
\usepackage{color}
\usepackage[dvipsnames]{xcolor}
\usepackage{amssymb}

\definecolor{light-gray}{gray}{0.78}
\definecolor{mid-gray}{gray}{0.55}
\definecolor{dark-gray}{gray}{0.32}

\begin{document}
\title{ Study muon g-2 at two-loop level in the $U(1)_X$SSM}
\author{Shu-Min Zhao$^{1,2}$\footnote{zhaosm@hbu.edu.cn}, Lu-Hao Su$^{1,2}$\footnote{suluhao0606@163.com}, Xing-Xing Dong$^{1,2}$\footnote{dxx$\_$0304@163.com}, Tong-Tong Wang$^{1,2}$,
Tai-Fu Feng$^{1,2,3}$\footnote{fentf@hbu.edu.cn}}

\affiliation{$^1$ Department of Physics, Hebei University, Baoding 071002, China}
\affiliation{$^2$ Key Laboratory of High-precision Computation and Application of Quantum Field Theory of Hebei Province, Baoding 071002, China}
\affiliation{$^3$ Department of Physics, Chongqing University, Chongqing 401331, China}
\date{\today}
\date{\today}
\begin{abstract}
The new experiment data of muon g-2 is reported by the workers at Fermilab National Accelerator Laboratory(FNAL).
Combined with the previous Brookhaven National Laboratory(BNL) E821 result, the departure from the standard model prediction is about 4.2 $\sigma$.
It strengthens our faith in the new physics.
$U(1)_X$SSM is the U(1) extension of the minimal supersymmetric standard model, where we study the electroweak corrections
to the anomalous magnetic dipole moment of muon from the one-loop diagrams and some two-loop diagrams possessing important contributions.
These two-loop diagrams include Barr-Zee type, rainbow type and diamond type. The virtual supersymmetric particles in these two-loop diagrams are
chargino, scalar neutrino, neutralino, scalar lepton, which are supposed not very heavy to make relatively large corrections.
We obtain the Wilson coefficients of the dimension 6 operators inducing the anomalous magnetic dipole moment of muon.
The numerical results can reach $25\times 10^{-10}$ and even larger.
\end{abstract}



\maketitle

\section{Introduction}

In the development of the standard model(SM) \cite{1948Phys}, the anomalous magnetic dipole moment (MDM) of muon has played a huge role.
The SM contributions to muon MDM include several parts:
The QED loop contributions $a_\mu^{QED}=116584718.931(104)\times 10^{-11}$ \cite{g2rep2020,add,GWB,AKDN1,GCMH,MHBL,MDAH,AKDN2,TBPA,TAMH,GCFH,GECS,TBNC,TATK,ACWJ,CGDS}; The electroweak contributions $a^{EW}_\mu=153.6(1.0)\times 10^{-11}$ \cite{ACWJ,CGDS};
The hadronic vacuum polarization contributions $a^{HVP}=6845(40)\times 10^{-11}$ \cite{g2rep2020,GCMH, had2}; The hadronic light-by-light contributions
$a^{HLBL}=92(18)\times 10^{-11}$ \cite{GCFH, GECS, TBNC}. Combining these results, one can obtain the SM prediction of muon anomaly
$a^{SM}_\mu=116591810(43)\times 10^{-11}$(0.37ppm) \cite{ g2rep2020, muon2, mdm2, TBPA}.

The MDM of muon $a_\mu\equiv(g_\mu-2)/2$  has been detected recently by the Fermilab National Accelerator Laboratory(FNAL)
muon g-2 experiment\cite{fnal,wx1,wx2,wx3,wx4,wx5,wx6,wx7}. The result is $a_{\mu}^{FNAL}=116592040(54)\times 10^{-11}$(0.46ppm) \cite{046} and 3.3 standard deviations larger than the SM prediction,
which is in great agreement with BNL E821 result\cite{GWB}. Then the new averaged experiment value of
 muon anomaly is $a^{exp}_{\mu}=116592061(41)\times 10^{-11}$(0.35ppm).
  The deviation between experiment and SM prediction is
$\Delta a_\mu=a^{exp}_\mu-a^{SM}_\mu=251(59)\times 10^{-11}$, which is 4.2$\sigma$.
The present deviation($4.2\sigma$) between the SM prediction and
the experimental data for $a_\mu$ is more stable, and it indicates that there should be new physics beyond the SM.
With the improvement of both experimental precision and theoretical prediction, the deviation will become more
important in the future, and open a window to explore new physics.

The electroweak  one-loop corrections from new physics
sector are generally suppressed by the factor $\Lambda_{{\rm EW}}^2/\Lambda^2$ with $\Lambda_{{\rm EW}}$ denoting the electroweak energy scale and
$\Lambda$ representing the new physics energy scale. Supposing
 the masses of neutralinos, charginos and scalar leptons of the second
generation equal to $M_{SUSY}$\cite{susy1, susy2}, the authors obtain the
approximated supersymmetric (SUSY) one-loop contributions by simplification
\begin{eqnarray}
|a_{\mu}^{SUSY}|=13\times10^{-10}\Big(
\frac{100{\rm GeV}} {M_{SUSY}}\Big)^2\tan\beta{\rm sign}(\mu_{H})
\label{susyoneloop}.
\end{eqnarray}
From this equation, one can estimate the one-loop SUSY contributions well. To obtain one-loop SUSY corrections around $2\times10^{-9}$, the parameters should
be in the region $M_{SUSY}\sim(200,500)$ GeV, $\tan\beta\sim(5,50)$. If $M_{SUSY}$ is larger than 500 GeV, the one-loop SUSY corrections are suppressed strongly.
The authors\cite{wx3} study the muon g-2 in
several GUT-scale constrained SUSY models including CMSSM/mSUGRA, pMSSM, CMSSM/mSUGRA extensions and GMSB/AMSB extensions.
In the general next-to-minimal supersymmetric
standard model (GNMSSM) with a singlino-dominated neutralino as a dark matter candidate, the numerical
result of muon g-2 is researched with the MultiNest technique for the parameter space\cite{caojj}.
To probe new physics accounting for muon g-2 and gravitational waves
with pulsar timing array measurements, the authors\cite{shuj} study the model possessing a light gauge boson or neutral scalar
interacting with muons.

The current experimental precision is high and sensitive to the two-loop electroweak corrections.
The one-loop correction to muon g-2 is well researched, but the two-loop study is more complicated and less advanced.
In some parameter space, the two-loop SUSY contributions become very important.
In the minimal supersymmetric extension of the standard model(MSSM), the contributions to muon MDM from
two-loop Barr-Zee type diagrams with sub-scalar-loop\cite{s1} and two-loop
diagrams enhanced by two powers of $\tan\beta$\cite{s2} are researched.
The photonic SUSY two-loop diagrams\cite{s3} where an additional photon loop is attached
to a SUSY one-loop diagram are studied in compact and analytic expressions.
There are non-decoupling two-loop corrections
from fermion/sfermion loops\cite{s4,s5} in the MSSM.
These corrections can be generally large
and even logarithmically enhanced for heavy sfermions.
The analytical results are presented and they obtain a very compact formula with approximation\cite{s5}.
GM2Calc (a public C++ program) is used to calculate precise MSSM prediction for muon MDM\cite{s6}.
In the CP-violating MSSM, there are some special two-loop diagrams belonging to diamond type,
which include virtual neutralino-slepton-Z\cite{two1}, chargino-sneutrino, neutralino-slepton, slepton-sneutrino\cite{two2},
and their corrections to muon MDM are studied.

 Employing an effective Lagrangian approach,
the authors\cite{other1} derive the leading-logarithm two-loop and three-loop
electroweak contributions to the muon MDM.
In Refs.\cite{other2,other3}, the authors research corrections
to muon MDM from the two-loop rainbow diagrams and Barr-Zee diagrams with heavy fermion sub-loop.
Furthermore, the analytic results are simplified at the decoupling limit and the leading corrections
are obtained obviously\cite{other2}.
The two-loop Barr-Zee type diagrams with  sub-fermion-loop and sub-scalar-loop between
 vector boson and Higgs are studied in BLMSSM\cite{one}.

The MSSM is one of the most popular models of new physics.
The authors present statistically convergent profile likelihood maps obtained via global
fits of a phenomenological MSSM with 15 free parameters, and analyze the phenomenology
of this model\cite{MSSM15fit}.
Including constraints from LHC data at 13 TeV and other experiments, a
frequentist analysis of the constraints on a phenomenological MSSM
model with 11 parameters is performed with MasterCode\cite{MSSM11fit}.
With the new GAMBIT software framework, the authors study a global fit of the MSSM\cite{MSSMgfit}.
Many people also study the phenomenology of MSSM including the anomaly of muon g-2, dark matter
relic density and direct detection, LHC constraints and so on. So, the viable parameter space of MSSM is
compressed\cite{MSSMg2DM,wx5,wx6,wx7} by more and more accurate experimental progresses.

Furthermore, the MSSM has $\mu$ problem, and  can not produce tiny mass to light neutrinos.
With these issues in mind, physicists extend MSSM. In these MSSM extensions, the U(1) extension is interesting.
We call the $U(1)_X$ extension of MSSM as $U(1)_X$SSM \cite{Sarah, ZSMJHEP}, where
three Higgs singlets and right-handed neutrinos are added to MSSM. Two gauge groups $U(1)_X$ and $U(1)_Y$ have gauge mixing. The right-handed neutrinos have two effects:
1. produce tiny mass to light neutrinos through see-saw mechanism, 2. provide a new dark matter candidate-light sneutrino.
This model relieves the so called little hierarchy problem that appears in the MSSM.
$\hat{S}$ is the singlet Higgs superfield with a non-zero VEV ($v_S/\sqrt{2}$).
 The terms $\mu\hat{H}_u\hat{H}_d$ and $\lambda_H\hat{S}\hat{H}_u\hat{H}_d$
 can produce an effective $\mu_{eff}=\mu+\lambda_Hv_S/\sqrt{2}$, which  relieves the $\mu$ problem.
Comparing with the condition in MSSM, the lightest CP-even Higgs mass at tree level is improved.
The second light neutral CP-even Higgs can be at TeV order. Then it easily satisfies the constraints for heavy Higgs from
experiments.

In our previous work, we calculate the contributions to muon MDM from some two-loop diagrams under the $U(1)_X$SSM with the effective Lagrangian method.
In this work, we research one-loop diagrams and more two-loop diagrams than our previous work \cite{slh}.
To make up the departure between experiment data and SM prediction for $a_\mu$, the scalar
neutrino and scalar lepton should not be heavy.
In the Fig.\ref{twolooptu}, we show the studied two-loop self-energy diagrams. $a_\mu$ is deduced from the triangle diagrams of the process
$\mu\rightarrow \mu+\gamma$.   Attaching a photon on the internal lines of the two-loop self-energy diagrams in all possible way, one
can get the two loop triangle diagrams. A two-loop self-energy diagram can produce several two-loop triangle diagrams, and the sum of
their amplitudes  satisfies Ward-identity. In this work,  we study the electroweak corrections
from several type two-loop SUSY diagrams(Barr-Zee type, rainbow type and diamond type) and the virtual SUSY particles include chargino, neutralino,
scalar lepton and scalar neutrino.
After tedious calculation and using on-shell condition for the external leptons, we get all dimension 6 operators and their coefficients. We neglect higher dimensional operators such
as dimension 8 operators, because they are tiny.

After this introduction, we show the main content of $U(1)_X$SSM and its superfields in section 2.
The one-loop and two-loop analytic results of $a_\mu$ are shown in the section 3.
The numerical results are shown in the section 4. The last section is used for the discussion and conclusion.

\section{The main content of $U(1)_X$SSM}
$U(1)_X$SSM is the U(1) extension of MSSM, whose local gauge group is $SU(3)_C\otimes
SU(2)_L \otimes U(1)_Y\otimes U(1)_X$. Comparing with MSSM, $U(1)_X$SSM has more superfields including:
right-handed neutrinos and three Higgs singlets. Through the see-saw mechanism, three light neutrinos obtain tiny masses at tree level.
The neutral CP-even Higgs mass squared matrix is $5\times5$, because of the introduction of $\eta,~\bar{\eta}$ and $S$. The lightest
CP-even Higgs mass can be improved at tree level. The particle contents can be found in our previous work\cite{ZSMJHEP,slh}.

The superpotential of $U(1)_X$SSM is
\begin{eqnarray}
&&W=l_W\hat{S}+\mu\hat{H}_u\hat{H}_d+M_S\hat{S}\hat{S}-Y_d\hat{d}\hat{q}\hat{H}_d-Y_e\hat{e}\hat{l}\hat{H}_d+\lambda_H\hat{S}\hat{H}_u\hat{H}_d
\nonumber\\&&+\lambda_C\hat{S}\hat{\eta}\hat{\bar{\eta}}+\frac{\kappa}{3}\hat{S}\hat{S}\hat{S}+Y_u\hat{u}\hat{q}\hat{H}_u+Y_X\hat{\nu}\hat{\bar{\eta}}\hat{\nu}
+Y_\nu\hat{\nu}\hat{l}\hat{H}_u.
\end{eqnarray}
We show the concrete forms of the  two Higgs doublets and three Higgs singlets
\begin{eqnarray}
&&H_{u}=\left(\begin{array}{c}H_{u}^+\\{1\over\sqrt{2}}\Big(v_{u}+H_{u}^0+iP_{u}^0\Big)\end{array}\right),
~~~~~~
H_{d}=\left(\begin{array}{c}{1\over\sqrt{2}}\Big(v_{d}+H_{d}^0+iP_{d}^0\Big)\\H_{d}^-\end{array}\right),
\nonumber\\
&&\eta={1\over\sqrt{2}}\Big(v_{\eta}+\phi_{\eta}^0+iP_{\eta}^0\Big),~~~~~~~~~~~~~~~
\bar{\eta}={1\over\sqrt{2}}\Big(v_{\bar{\eta}}+\phi_{\bar{\eta}}^0+iP_{\bar{\eta}}^0\Big),\nonumber\\&&
\hspace{4.0cm}S={1\over\sqrt{2}}\Big(v_{S}+\phi_{S}^0+iP_{S}^0\Big).
\end{eqnarray}
The VEVs of the Higgs superfields $H_u$, $H_d$, $\eta$, $\bar{\eta}$ and $S$ are presented by
$v_u,~v_d,~v_\eta$,~ $v_{\bar\eta}$ and $v_S$ respectively. Two angles are defined as
 $\tan\beta=v_u/v_d$ and $\tan\beta_\eta=v_{\bar{\eta}}/v_{\eta}$.

The soft SUSY breaking terms are
\begin{eqnarray}
&&\mathcal{L}_{soft}=\mathcal{L}_{soft}^{MSSM}-B_SS^2-L_SS-\frac{T_\kappa}{3}S^3-T_{\lambda_C}S\eta\bar{\eta}
+\epsilon_{ij}T_{\lambda_H}SH_d^iH_u^j\nonumber\\&&
-T_X^{IJ}\bar{\eta}\tilde{\nu}_R^{*I}\tilde{\nu}_R^{*J}
+\epsilon_{ij}T^{IJ}_{\nu}H_u^i\tilde{\nu}_R^{I*}\tilde{l}_j^J
-m_{\eta}^2|\eta|^2-m_{\bar{\eta}}^2|\bar{\eta}|^2\nonumber\\&&
-m_S^2S^2-(m_{\tilde{\nu}_R}^2)^{IJ}\tilde{\nu}_R^{I*}\tilde{\nu}_R^{J}
-\frac{1}{2}\Big(M_S\lambda^2_{\tilde{X}}+2M_{BB^\prime}\lambda_{\tilde{B}}\lambda_{\tilde{X}}\Big)+h.c~~.
\end{eqnarray}
 $Y^Y$ denotes the $U(1)_Y$ charge and $Y^X$ represents the $U(1)_X$ charge.
 We have proven that $U(1)_X$SSM is anomaly free.
 Two Abelian groups $U(1)_Y$ and $U(1)_X$ in $U(1)_X$SSM produce a new effect: the gauge kinetic mixing, which
can also be induced through RGEs even with zero value at $M_{GUT}$.

In the general form, the covariant derivatives of $U(1)_X$SSM reads as \cite{model1, model2, model3}
\begin{eqnarray}
&&D_\mu=\partial_\mu-i\left(\begin{array}{cc}Y^Y,&Y^X\end{array}\right)
\left(\begin{array}{cc}g_{Y},&g{'}_{{YX}}\\g{'}_{{XY}},&g{'}_{{X}}\end{array}\right)
\left(\begin{array}{c}A_{\mu}^{\prime Y} \\ A_{\mu}^{\prime X}\end{array}\right)\;.
\label{gauge1}
\end{eqnarray}
The gauge fields of $U(1)_Y$ and $U(1)_X$ are denoted by
$A_{\mu}^{\prime Y}$ and $A^{\prime X}_\mu$. With the two Abelian gauge groups unbroken condition, we use the
matrix $R$ \cite{model1, model3} to obtain
\begin{eqnarray}
&&\left(\begin{array}{cc}g_{Y},&g{'}_{{YX}}\\g{'}_{{XY}},&g{'}_{{X}}\end{array}\right)
R^T=\left(\begin{array}{cc}g_{1},&g_{{YX}}\\0,&g_{{X}}\end{array}\right)\;.
\label{gauge2}
\end{eqnarray}

In this model, the gauge bosons $A^{X}_\mu,~A^Y_\mu$ and $V^3_\mu$ mix together at the tree level. We show the mass matrix
in the basis $(A^Y_\mu, V^3_\mu, A^{X}_\mu)$ as
\begin{eqnarray}
&&\left(\begin{array}{*{20}{c}}
\frac{1}{8}g_{1}^2 v^2 &~~~ -\frac{1}{8}g_{1}g_{2} v^2 & ~~~\frac{1}{8}g_{1}(g_{YX}+g_X) v^2 \\
-\frac{1}{8}g_{1}g_{2} v^2 &~~~ \frac{1}{8}g_{2}^2 v^2 & ~~~~-\frac{1}{8}g_{2}(g_{YX}+g_X) v^2\\
\frac{1}{8}g_{1}(g_{YX}+g_X) v^2 &~~~ -\frac{1}{8}g_{2}(g_{YX}+g_X) v^2 &~~~~ \frac{1}{8}(g_{YX}+g_X)^2 v^2+\frac{1}{8}g_{{X}}^2 \xi^2
\end{array}\right),\label{gauge matrix}
\end{eqnarray}
with $v^2=v_u^2+v_d^2$ and $\xi^2=v_\eta^2+v_{\bar{\eta}}^2$.
 Two mixing angles $\theta_{W}$ and $\theta_{W}'$ are used to diagonalize the mass matrix in Eq. (\ref{gauge matrix}).
 $\sin^2\theta_{W}^\prime$ is defined as
\begin{eqnarray}
\sin^2\theta_{W}'=\frac{1}{2}-\frac{((g_{YX}+g_X)^2-g_{1}^2-g_{2}^2)v^2+
4g_{X}^2\xi^2}{2\sqrt{((g_{YX}+g_X)^2+g_{1}^2+g_{2}^2)^2v^4+8g_{X}^2((g_{YX}+g_X)^2-g_{1}^2-g_{2}^2)v^2\xi^2+16g_{X}^4\xi^4}}.
\end{eqnarray}
We deduce the eigenvalues of Eq. (\ref{gauge matrix})
\begin{eqnarray}
&&m_\gamma^2=0,\nonumber\\
&&m_{Z,{Z^{'}}}^2=\frac{1}{8}\Big((g_{1}^2+g_2^2+(g_{YX}+g_X)^2)v^2+4g_{X}^2\xi^2 \nonumber\\
&&\mp\sqrt{(g_{1}^2+g_{2}^2+(g_{YX}+g_X)^2)^2v^4+8((g_{YX}+g_X)^2-g_{1}^2-
g_{2}^2)g_{X}^2v^2\xi^2+16g_{X}^4\xi^4}\Big).
\end{eqnarray}
These results are similar as the condition of B-LSSM with the relation $g_{YX}+g_X\rightarrow g_{YX}$.
The reason comes from the differences of the $Y^X$ charges of the superfields in the both models.
The used mass matrixes can be found in the work \cite{ffa}. Here, we show some used couplings.

We also deduce the vertexes of $\bar{l}_i-\chi_j^--\tilde{\nu}^R_k(\tilde{\nu}^I_k)$
\begin{eqnarray}
&&\mathcal{L}_{\bar{l}\chi^-\tilde{\nu}^R}=\frac{1}{\sqrt{2}}\bar{l}_i\Big\{U^*_{j2}Z^{R*}_{ki}Y_l^iP_L
-g_2V_{j1}Z^{R*}_{ki}P_R\Big\}\chi_j^-\tilde{\nu}^R_k,\nonumber\\
&&\mathcal{L}_{\bar{l}\chi^-\tilde{\nu}^I}=\frac{i}{\sqrt{2}}\bar{l}_i\Big\{U^*_{j2}Z^{I*}_{ki}Y_l^iP_L
-g_2V_{j1}Z^{I*}_{ki}P_R\Big\}\chi_j^-\tilde{\nu}^I_k.
\end{eqnarray}

The vertexes of $\bar{\chi}_i^0-\nu_i-\tilde{\nu}^R_k(\tilde{\nu}^I_k)$ are
\begin{eqnarray}
&&\mathcal{L}_{\bar{\chi}^0\nu\tilde{\nu}^R}=\frac{1}{2}\bar{\chi}_i^0\Big\{(-g_2N^*_{i2}+g_{YX}N^*_{i5}+g_1N^*_{i1})
\sum_{a=1}^3Z^{R*}_{ka}U_{ja}^{V*}P_L\nonumber\\&&\hspace{1.7cm}+
(-g_2N_{i2}+g_{YX}N_{i5}+g_1N_{i1})\sum_{a=1}^3Z^{R*}_{ka}U_{ja}^{V}P_R\Big\}\nu_i\tilde{\nu}^R_k,\nonumber
\\&&\mathcal{L}_{\bar{\chi}^0\nu\tilde{\nu}^I}=-\frac{i}{2}\bar{\chi}_i^0\Big\{(-g_2N^*_{i2}+g_{YX}N^*_{i5}+g_1N^*_{i1})
\sum_{a=1}^3Z^{I*}_{ka}U_{ja}^{V*}P_L\nonumber\\&&\hspace{1.7cm}+
(g_2N_{i2}-g_{YX}N_{i5}-g_1N_{i1})\sum_{a=1}^3Z^{I*}_{ka}U_{ja}^{V}P_R\Big\}\nu_i\tilde{\nu}^I_k.
\end{eqnarray}
The vertexes of W-slepton-sneutrino(CP-even and CP-odd) read as
\begin{eqnarray}
&&\mathcal{L}_{\tilde{L}\tilde{\nu}^{R*}W}=-\frac{1}{2}g_2\tilde{L}_i\tilde{\nu}^{R*}_j
\sum_{a=1}^3Z^{E*}_{ia}Z^{R*}_{ja}(-p_{\mu}^{\tilde{\nu}_j^R}+p_\mu^{\tilde{L}_i})W^\mu
+h.c,\nonumber\\&&
\mathcal{L}_{\tilde{L}\tilde{\nu}^{I*}W}=\frac{i}{2}g_2\tilde{L}_i\tilde{\nu}^{I*}_j
\sum_{a=1}^3Z^{E*}_{ia}Z^{I*}_{ja}(-p_{\mu}^{\tilde{\nu}_j^I}+p_\mu^{\tilde{L}_i})W^\mu
+h.c.
\end{eqnarray}

We also deduce the vertex couplings of neutrino-slepton-chargino and neutralino-lepton-slepton
\begin{eqnarray}
&&\mathcal{L}_{\bar{\nu}\chi^-\tilde{L}}=\bar{\nu}_i\Big((-g_2U^*_{j1}\sum_{a=1}^3U^{V*}_{ia}Z^E_{ka}+U^*_{j2}\sum_{a=1}^3U^{V*}_{ia}Y^a_lZ^E_{k(3+a)})P_L
\nonumber\\&&\hspace{1.6cm}+\sum_{a,b=1}^3Y_{\nu}^{ab}U^V_{i(3+a)}Z^E_{kb}V_{j2}P_R\Big)\chi^-_j\tilde{L}_k,
\\
&&\mathcal{L}_{\bar{\chi}^0l\tilde{L}}=\bar{\chi}^0_i\Big\{\Big(\frac{1}{\sqrt{2}}(g_1N^*_{i1}+g_2N^*_{i2}+g_{YX}N^*_{i5})Z^E_{kj}
-N^*_{i3}Y^j_lZ^E_{k(3+j)}\Big)P_L\nonumber\\&&\hspace{1.6cm}
-\Big[\frac{1}{\sqrt{2}}\Big(2g_1N_{i1}+(2g_{YX}+g_X)N_{i5}\Big)Z^E_{k(3+a)}+Y_{l}^jZ^E_{kj}N_{i3}\Big]P_R\Big\}l_j\tilde{L}_k.
\end{eqnarray}
The other needed couplings can be found in our previous works\cite{ZSMJHEP,slh}.

\section{formulation}
  For the process $l^I\rightarrow l^I+\gamma$, the Feynman amplitude can be expressed by these dimension 6 operators \cite{lepton} with the effective Lagrangian method.
 For muon MDM, the dimension 8 operators are suppressed by additional factor $\frac{m_{\mu}^2}{M_{SUSY}^2}$ $\sim$ ($10^{-7}$, $10^{-8}$) and are neglected safely.
 These dimension 6 operators are shown as
\begin{eqnarray}
&&\mathcal{O}_1^{L,R}=\frac{1}{(4\pi)^2}\bar{l}(i\mathcal{D}\!\!\!\slash)^3P_{L,R}l,~~~~~~~~~~~~~~
\mathcal{O}_2^{L,R}=\frac{eQ_f}{(4\pi)^2}\overline{(i\mathcal{D}_{\mu}l)}\gamma^{\mu}
F\cdot\sigma P_{L,R}l,
\nonumber\\
&&\mathcal{O}_3^{L,R}=\frac{eQ_f}{(4\pi)^2}\bar{l}F\cdot\sigma\gamma^{\mu}
P_{L,R} (i\mathcal{D}_{\mu}l),~~~~\mathcal{O}_4^{L,R}=\frac{eQ_f}{(4\pi)^2}\bar{l}(\partial^{\mu}F_{\mu\nu})\gamma^{\nu}
P_{L,R}l,\nonumber\\&&
\mathcal{O}_5^{L,R}=\frac{m_l}{(4\pi)^2}\bar{l}(i\mathcal{D}\!\!\!\slash)^2P_{L,R}l,
~~~~~~~~~~~~~~\mathcal{O}_6^{L,R}=\frac{eQ_fm_l}{(4\pi)^2}\bar{l}F\cdot\sigma
P_{L,R}l,
\end{eqnarray}
with $\mathcal{D}_{\mu}=\partial_{\mu}+ieA_{\mu}$ and $P_{L,R}=\frac{1\mp\gamma_5}{2}$. $F_{{\mu\nu}}$ is the electromagnetic field strength, and
$m_{l}$ is the lepton mass. The operators $\mathcal{O}_{2,3,6}^{L,R}$ have relation with lepton MDM, which is the combination of the Wilson coefficients $C^{L,R}_{2,3,6}$.
 Using the on-shell condition for the external leptons, one can obtain lepton MDM from the following effective Lagrangian.
\begin{eqnarray}
&&{\cal L}_{{MDM}}={e\over4m_{l}}\;a_{l}\;\bar{l}\sigma^{\mu\nu}
l\;F_{{\mu\nu}}\label{adm}.
\end{eqnarray}

\subsection{The one-loop corrections}

\begin{figure}[h]
\setlength{\unitlength}{1mm}
\centering
\includegraphics[width=3.6in]{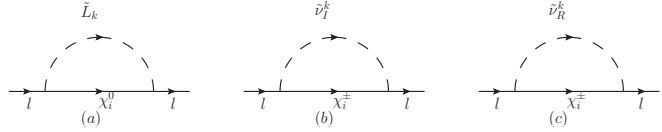}
\caption[]{ The one-loop self-energy diagrams \label{onelooptu}}
\end{figure}

In $U(1)_X$SSM, the one-loop self-energy diagrams of lepton are shown in the Fig.\ref{onelooptu}. Attaching a photon on the
internal lines of the one-loop self-energy diagram in all possible way, one can obtain the triangle diagrams for $l\rightarrow l+\gamma$.

The analytic forms of the one-loop contributions are collected here.

1. The corrections from neutralinos and scalar leptons
\begin{eqnarray}
&&a_{\mu}^{1L,~\tilde{L}\chi^{0}}=
-\sum_{k=1}^6\sum_{j=1}^8\Big[\Re(A_L^*A_R)
\sqrt{x_{\chi_j^{0}}x_{\mu}}x_{\tilde{L}_k}\frac{\partial^2 \mathcal{B}(x_{\chi_j^{0}},x_{\tilde{L}_k})}{\partial x_{\tilde{L}_k}^2}
\nonumber\\&&\hspace{1.4cm}+\frac{1}{3}(|A_L|^2+|A_R|^2)x_{\tilde{L}_k}x_{\mu}
\frac{\partial\mathcal{B}_1(x_{\chi_j^{0}},x_{\tilde{L}_k})}{\partial x_{\tilde{L}_k}}\Big].
\end{eqnarray}
with $x=\frac{m^2}{\Lambda^2}$. Here, $m$ is the particle mass.
The concrete forms of  $A_R$ and $A_L$ read as
\begin{eqnarray}
&&A_R=\frac{1}{\sqrt{2}}Z_{k2}^{E}(g_1N_{j1}^{*}+g_2N_{j2}^{*}+g_{YX}N_{j5}^{*})
-N_{j3}^{*}Y_\mu Z_{k5}^{E},\nonumber\\&&
A_L=-\frac{1}{\sqrt{2}}Z_{k5}^{E}[2g_1N_{j1}+(2g_{YX}+g_X)N_{j5}]-Y_\mu^{*}Z_{k2}^EN_{j3}.
\end{eqnarray}

We show the functions $\mathcal{B}(x,y)$ and $\mathcal{B}_1(x,y)$ explicitly
\begin{eqnarray}
\mathcal{B}(x,y)=\frac{1}{16 \pi
^2}\Big(\frac{x \ln x}{y-x}+\frac{y \ln
y}{x-y}\Big),~~~
\mathcal{B}_1(x,y)=(
\frac{\partial}{\partial y}+\frac{y}{2}\frac{\partial^2 }{\partial y^2})\mathcal{B}(x,y).
\end{eqnarray}

2. The corrections from chargino and  CP-odd scalar neutrino
\begin{eqnarray}
&&a_{\mu}^{1L,~\tilde{\nu}^I\chi^{\pm}}=\sum_{i=1}^2\sum_{k=1}^6
\Big[-2\Re(B_L^{*}B_R)\sqrt{x_{\chi_i^{-}}x_\mu}\mathcal{B}_1(x_{\tilde{\nu}^I_k},x_{\chi_i^{-}})
\nonumber\\&&\hspace{1.8cm}+\frac{1}{3}(|B_L|^2+|B_R|^2)x_\mu x_{\chi_i^{-}}\frac{\partial\mathcal{B}_1(x_{\tilde{\nu}^I_k},x_{\chi_i^{-}})}{\partial x_{\chi_i^{-}}}\Big].
\nonumber\\&&
B_L=-\frac{1}{\sqrt{2}}U_{i2}^{*}Z_{k2}^{I*}Y_\mu,~~~
B_R=\frac{1}{\sqrt{2}}g_2Z_{k2}^{I*}V_{i1}.
\end{eqnarray}

3. The corrections from chargino and  CP-even scalar neutrino
\begin{eqnarray}
&&a_{\mu}^{1L,~\tilde{\nu}^R\chi^{\pm}}=\sum_{i=1}^2\sum_{k=1}^6
\Big[-2\Re(C_L^{*}C_R)\sqrt{x_{\chi_i^{-}}x_\mu}\mathcal{B}_1(x_{\tilde{\nu}^R_k},x_{\chi_i^{-}})
\nonumber\\&&\hspace{1.8cm}+\frac{1}{3}(|C_L|^2+|C_R|^2)x_\mu x_{\chi_i^{-}}\frac{\partial\mathcal{B}_1(x_{\tilde{\nu}^R_k},x_{\chi_i^{-}})}{\partial x_{\chi_i^{-}}}\Big].
\nonumber\\&&
C_L=\frac{1}{\sqrt{2}}U_{i2}^{*}Z_{k2}^{R*}Y_\mu,~~~
C_R=-\frac{1}{\sqrt{2}}g_2Z_{k2}^{R*}V_{i1}.
\end{eqnarray}

4. The corrections from the new vector boson $Z^\prime$ and lepton. The mass of $Z^\prime$ are very heavy, and we take $m_{Z^\prime}$ larger than 5.1 TeV.
Comparing with Z-lepton one-loop contribution, the corresponding contributions from $Z^\prime$-lepton are suppressed by the factor $\frac{m_Z^2}{m^2_{Z^\prime}}\sim 4\times 10^{-4}$.
Therefore, we neglect $Z^\prime$-lepton one-loop contribution.

5. The neutral Higgs-lepton and charged Higgs-neutrino contributions are suppressed by the square of the Higgs-lepton coupling $\frac{m_{\mu}^2}{m^2_W}\sim10^{-6}$.
As discussed in Ref.\cite{our},
these type contributions are neglected.

The one-loop contributions to muon g-2 can be expressed as
\begin{eqnarray}
a_\mu^{1L}=a_{\mu}^{1L,~\tilde{L}\chi^{0}}+a_{\mu}^{1L,~\tilde{\nu}^R\chi^{\pm}}+a_{\mu}^{1L,~\tilde{\nu}^I\chi^{\pm}}.
\end{eqnarray}
For the one-loop contributions, in fact the factor does not represent an enhancement
proportional to $m_\chi/m_\mu$, because it is suppressed by the combined rotation matrixes.
In the end, they produce an overall enhancement factor $\tan\beta$\cite{dabeta1,dabeta2}. The similar condition is also
for the two-loop contributions. The apparent factor $m_\chi/m_\mu$ in the two-loop corrections is
also suppressed by the combined rotation matrixes and can not yield enhancement
proportional to $m_\chi/m_\mu$ in the total analysis.
To show the factor more clearly, we give the one-loop corrections with
the mass insertion approximation (MIA)\cite{wx1,wx7,dabeta1,MIAhao}.
These approximations indeed
clarify the major parameter dependence (the appearance of $\tan\beta$ instead of $m_\chi/m_\mu$).
Here, using MIA we obtain the concrete forms of the one-loop g-2 results in $U(1)_X$SSM.

1. The one-loop contributions from chargino and CP-even(odd) sneutrino.
\begin{eqnarray}
&&a_\mu(\tilde{\nu}^R_L, \tilde{H}^-, \tilde{W}^-)
=\frac{g_2^2}{2}
x_\mu\sqrt{x_2x_{\mu^{\prime}_H}}\tan\beta[2\mathcal{I}(x_{\mu^{\prime}_H},x_{\tilde{\nu}^R_L},x_2)
-\mathcal{J}(x_2,x_{\mu^{\prime}_H},x_{\tilde{\nu}^R_L})
\nonumber\\&&\hspace{3.8cm}+2\mathcal{I}(x_2,x_{\tilde{\nu}^R_L},x_{\mu^{\prime}_H})
-\mathcal{J}(x_{\mu^{\prime}_H},x_2,x_{\tilde{\nu}^R_L})]\label{MIARC},
\\&&a_\mu(\tilde{\nu}^I_L, \tilde{H}^-, \tilde{W}^-)
=\frac{g_2^2}{2}
x_\mu\sqrt{x_2x_{\mu^{\prime}_H}}\tan\beta[2\mathcal{I}(x_{\mu^{\prime}_H},x_{\tilde{\nu}^I_L},x_2)
-\mathcal{J}(x_2,x_{\mu^{\prime}_H},x_{\tilde{\nu}^I_L})
\nonumber\\&&\hspace{3.8cm}+2\mathcal{I}(x_2,x_{\tilde{\nu}^I_L},x_{\mu^{\prime}_H})
-\mathcal{J}(x_{\mu^{\prime}_H},x_2,x_{\tilde{\nu}^I_L})]\label{MIAIC}.
\end{eqnarray}
with $\mu_{H}^\prime=\frac{\lambda_H v_S}{\sqrt{2}}+\mu$
and $x_{\mu^{\prime}_H}=\frac{\mu_{H}^{\prime2}}{\Lambda^2}$.

The one-loop functions $\mathcal{I}(x,y,z)$ and $\mathcal{J}(x,y,z)$ read as
\begin{eqnarray}
&&\mathcal{J}(x,y,z)=\frac{1}{16\pi^2}\Big[\frac{x (x^2+x z-2 y
   z)\log x}{(x-y)^2 (x-z)^3}-\frac{y^2 \log
   y}{(x-y)^2 (y-z)^2}\nonumber\\&&\hspace{1.8cm}+\frac{z[x
   (z-2 y)+z^2] \log z}{(z-x)^3 (y-z)^2}-\frac{x
   (y-2 z)+y z}{(x-y) (x-z)^2 (y-z)}\Big].
\\&&\mathcal{I}(x,y,z)=\frac{1}{16\pi^2}\Big[\frac{1}{(x-z) (z-y)}+\frac{(z^2-x y)\log z}{(x-z)^2
   (y-z)^2}\nonumber\\&&\hspace{1.8cm}-\frac{x \log
   x}{(x-y) (x-z)^2}+\frac{y \log y}{(x-y)
   (y-z)^2}\Big].
\end{eqnarray}

2. The one-loop contributions from $\tilde{B}(\lambda_{\tilde{X}})$-$\tilde{\mu}_L$-$\tilde{\mu}_R$.
\begin{eqnarray}
&&a_\mu(\tilde{\mu}_R,\tilde{\mu}_L, \tilde{B})=
g_1^2x_\mu\sqrt{x_1x_{\mu^{\prime}_H}}
\tan\beta[\mathcal{J}(x_1,x_{\tilde{\mu}_L},x_{\tilde{\mu}_R})
+\mathcal{J}(x_1,x_{\tilde{\mu}_R},x_{\tilde{\mu}_L})]\label{MIABLR}
.\\
&&a_\mu(\tilde{\mu}_R,\tilde{\mu}_L, \lambda_{\tilde{X}})
=(g_{YX}^2
+\frac{g_Xg_{YX}}{2})x_\mu\sqrt{x_{\lambda_{\tilde{X}}}x_{\mu^{\prime}_H}}\tan\beta
\nonumber\\&&\hspace{3.0cm}\times[\mathcal{J}(x_{\lambda_{\tilde{X}}}^2,x_{\tilde{\mu}_L},x_{\tilde{\mu}_R})
+\mathcal{J}(x_{\lambda_{\tilde{X}}}^2,x_{\tilde{\mu}_R},x_{\tilde{\mu}_L})]\label{MIAXLR}.
\end{eqnarray}

3. The one-loop contributions from $\tilde{B}(\lambda_{\tilde{X}})$-$\tilde{H}^0$-$\tilde{\mu}_R$.
\begin{eqnarray}
&&a_\mu(\tilde{\mu}_R, \tilde{B}, \tilde{H}^0)
=-g_1^2x_\mu\sqrt{x_1x_{\mu^{\prime}_H}}\tan\beta[
\mathcal{J}(x_1^2,x_{\mu^{\prime}_H},x_{\tilde{\mu}_R})+\mathcal{J}(x_{\mu^{\prime}_H},x_1^2,x_{\tilde{\mu}_R})]\label{MIAHBR},
\\&&a_\mu(\tilde{\mu}_R, \lambda_{\tilde{X}}, \tilde{H}^0)
=-(g_{YX}+\frac{g_X}{2})(g_{YX}+g_X)x_\mu\sqrt{x_{\lambda_{\tilde{X}}}x_{\mu^{\prime}_H}}\tan\beta
\nonumber\\&&\hspace{3.2cm}\times[\mathcal{J}(x_{\lambda_{\tilde{X}}},x_{\mu^{\prime}_H},x_{\tilde{\mu}_R})
+\mathcal{J}(x_{\mu^{\prime}_H},x_{\lambda_{\tilde{X}}},x_{\tilde{\mu}_R})]\label{MIAHXR}.
\end{eqnarray}

4. The one-loop contributions from $\tilde{B}(\tilde{W}^0,\lambda_{\tilde{X}})$-$\tilde{H}^0$-$\tilde{\mu}_L$.
\begin{eqnarray}
&&a_\mu(\tilde{\mu}_L, \tilde{H}^0, \tilde{B})
=\frac{1}{2}g_1^2x_\mu\sqrt{x_1x_{\mu^{\prime}_H}}\tan\beta[
\mathcal{J}(x_1,x_{\mu^{\prime}_H},x_{\tilde{\mu}_L})+\mathcal{J}(x_{\mu^{\prime}_H},x_1,x_{\tilde{\mu}_L})]\label{MIABHL},
\\&&a_\mu(\tilde{\mu}_L, \tilde{H}^0, \tilde{W}^0)=-\frac{1}{2}g_2^2
x_\mu\sqrt{x_2x_{\mu^{\prime}_H}}\tan\beta[
\mathcal{J}(x_2,x_{\mu^{\prime}_H},x_{\tilde{\mu}_L})+\mathcal{J}(x_{\mu^{\prime}_H},x_2,x_{\tilde{\mu}_L})]\label{MIAWHL},
\\&&a_\mu(\tilde{\mu}_L, \tilde{H}^0, \lambda_{\tilde{X}})
=\frac{1}{2}(g_{YX}+g_X)g_{YX}x_\mu\sqrt{x_{\lambda_{\tilde{X}}}x_{\mu^{\prime}_H}}\tan\beta\nonumber\\&&\hspace{3cm}\times
[\mathcal{J}(x_{\lambda_{\tilde{X}}},x_{\mu^{\prime}_H},x_{\tilde{\mu}_L})+
\mathcal{J}(x_{\mu^{\prime}_H},x_{\lambda_{\tilde{X}}},x_{\tilde{\mu}_L})]\label{MIAXHL}.
\end{eqnarray}

5. The one-loop contributions from $\tilde{B}-\lambda_{\tilde{X}}-\tilde{\mu}_R-\tilde{\mu}_L$.
\begin{eqnarray}
&&a_\mu(\tilde{\mu}_R,\tilde{\mu}_L, \tilde{B}, \lambda_{\tilde{X}})
=g_1(4g_{YX}+g_X)x_\mu \sqrt{x_{BB^\prime}x_{\mu^\prime_H}}\tan\beta
\nonumber\\&&\times\Big(\sqrt{x_1x_{\lambda_{\tilde{X}}}}
f(x_{\lambda_{\tilde{X}}},x_1,x_{\tilde{\mu}_L},x_{\tilde{\mu}_R}) -g(x_{\lambda_{\tilde{X}}},x_1,x_{\tilde{\mu}_L},x_{\tilde{\mu}_R})
\Big)\label{MIAXBLR}.
\end{eqnarray}
We show the one loop functions $f(x,y,z,t)$ and $g(x,y,z,t)$ in the following form
\begin{eqnarray}
&&f(x,y,z,t)=
   \frac{1}{16\pi^2}\Big[\frac{t [t^3-3 t x y+x y
   (x+y)]\log t}{(t-x)^3 (t-y)^3 (t-z)}-\frac{x[x^3-3 t x z+t z (t+z)] \log x
}{(t-x)^3 (x-y)
   (x-z)^3}\nonumber\\&&+\frac{y [y^3-3 t y z+t z
   (t+z)]\log y}{(t-y)^3 (x-y) (y-z)^3}-\frac{z[z^3-3 x y z+x y (x+y)] \log z}{(t-z) (z-x)^3
   (z-y)^3}+\frac{1}{2
   (x-y)}\nonumber\\&&\times\Big(\frac{t}{(t-x)^2 (z-x)}-\frac{2y}{(t-y) (y-z)^2}+\frac{x (2 t-3
   x+z)}{(t-x)^2 (x-z)^2}+\frac{t+y}{(t-y)^2 (y-z)}\Big)\Big],
\\&&g(x,y,z,t)=\frac{1}{16\pi^2}\Big\{-\frac{t [t^3 (x+y)-3 t^2 x y+x^2
   y^2]\log t }{(t-x)^3 (t-y)^3 (t-z)}\nonumber\\&&+\frac{z [x^2 y^2+x z^2 (z-3 y)+y z^3]\log z
   }{(t-z) (z-x)^3
   (z-y)^3}+\frac{x^2[x^3-3 t x z+t z
   (t+z)] \log x}{(t-x)^3 (x-y) (x-z)^3}\nonumber\\&&-\frac{y^2 [y^3-3 t y z+t z (t+z)]\log y }{(t-y)^3 (x-y)
   (y-z)^3}-\frac{x^2 (2 t-3
   x+z)}{2(t-x)^2 (x-y) (x-z)^2}\nonumber\\&&+\frac{t x}{2(t-x)^2 (x-y)
   (x-z)}-\frac{y [t (y+z)+y (z-3 y)]}{2(t-y)^2 (y-x)
   (y-z)^2}\Big\}.
\end{eqnarray}

In Eqs.(\ref{MIARC}), (\ref{MIAIC}), (\ref{MIABLR}) $\dots$ (\ref{MIAXBLR}),
 one can easily find the factors $x_\mu=\frac{m_\mu^2}{\Lambda^2}$ and $\tan\beta$. This
 characteristic is same as the condition in MSSM.
 The contributions relating with the new gaugino $\lambda_{\tilde{X}}$ are shown in
 Eqs.(\ref{MIAXLR}), (\ref{MIAHXR}), (\ref{MIAXHL}), (\ref{MIAXBLR}),
 which include the new gauge coupling constants $g_X$ and $g_{YX}$.

 To obtain clearer images of the results, we suppose that
 all the masses of the superparticles are almost degenerate.
 The author\cite{dabeta1} gives the one-loop MSSM results(chargino-sneutrino, neutralino-slepton) in the extreme case
where the masses for superparticles$(M_1, M_2,
\mu_H, m_{\tilde{\mu}_L}, m_{\tilde{\mu}_R})$ are equal to $M_{SUSY}$
\begin{eqnarray}
a^{MSSM}_{\mu}\simeq\frac{1}{192\pi^2}\frac{m_\mu^2}{M_{SUSY}^2}\tan\beta(5g_2^2+g_1^2).\label{amumssm}
\end{eqnarray}

Here, we also use the similar case
\[M_1=M_2=\mu_H^\prime=m_{\tilde{\mu}_L}
=m_{\tilde{\mu}_R}=|M_{\lambda_{\tilde{X}}}|=|M_{BB^\prime}|=M_{SUSY},\]
and the four functions $\mathcal{I}(x,y,z),~\mathcal{J}(x,y,z),~f(x,y,z,t),~g(x,y,z,t)$ are much simplified as
\begin{eqnarray}
&&\mathcal{J}(1,1,1)=\frac{1}{192\pi^2},~~~~~~~~
 \mathcal{I}(1,1,1)= \frac{1}{96\pi^2},\nonumber\\&&
 f(1,1,1,1)= -\frac{1}{240 \pi^2 },~~~~g(1,1,1,1)= -\frac{1}{960 \pi ^2}.
\end{eqnarray}
In this condition, we obtain the much simplified one-loop results of muon g-2 in $U(1)_X$SSM.
\begin{eqnarray}
&&a^{1L}_\mu\simeq \frac{1}{192\pi^2}\frac{m_\mu^2}{M_{SUSY}^2}\tan\beta(5g_2^2+g_1^2)\nonumber\\&&
+\frac{1}{960\pi^2}\frac{m_\mu^2}{M_{SUSY}^2}\tan\beta\Big[5(g_{YX}^2-g_{YX}g_X-g_X^2)\texttt{sign}[M_{\lambda_{\tilde{X}}}]
\nonumber\\&&+g_1(4g_{YX}+g_X)\texttt{sign}[M_{BB^\prime}]
\Big(1-4\texttt{sign}[M_{\lambda_{\tilde{X}}}]\Big)\Big].\label{amuS}
\end{eqnarray}
The terms in the first line of Eq.(\ref{amuS}) is equal to the MSSM results in Eq.(\ref{amumssm}).
From this equation, we can see the new gaugino $\lambda_{\tilde{X}}$ can give considerable corrections to $a_\mu$.
In the  condition $1>g_X>g_{YX}>0$ and with the supposition $\texttt{sign}[M_{\lambda_{\tilde{X}}}]=-1$ and
$\texttt{sign}[M_{BB^\prime}]=1$, the corrections beyond MSSM can reach large value.
\begin{eqnarray}
&&a^{1L}_\mu\rightarrow\frac{1}{192\pi^2}\frac{m_\mu^2}{M_{SUSY}^2}\tan\beta\Big[(5g_2^2+g_1^2)
+(g_{YX}g_X+g_X^2 -g_{YX}^2+4g_1g_{YX}+g_1g_X)\Big].
\end{eqnarray}
Here, the order analysis shows
\begin{eqnarray}
0<\frac{g_{YX}g_X+g_X^2 -g_{YX}^2+4g_1g_{YX}+g_1g_X}{5g_2^2+g_1^2}\lesssim 1.
\end{eqnarray}
Then the $U(1)_X$SSM contributions beyond MSSM are considerable.

\subsection{The two-loop corrections}

\begin{figure}[h]
\setlength{\unitlength}{1mm}
\centering
\includegraphics[width=4.6in]{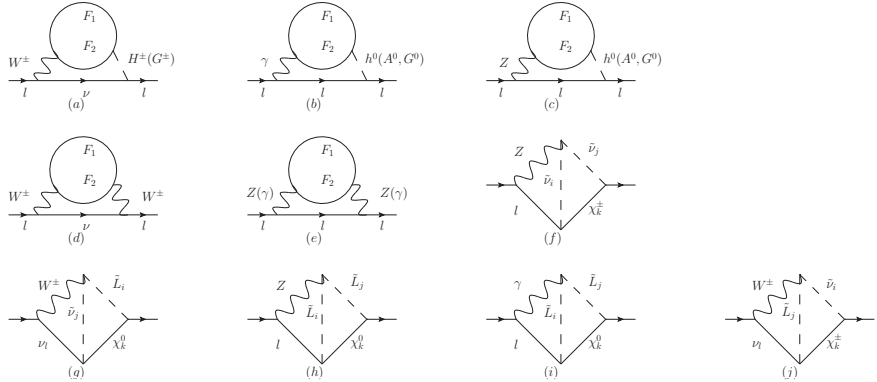}
\caption[]{ The two-loop self-energy diagrams \label{twolooptu}}
\end{figure}

In this work, we study major two-loop diagram contributions to muon MDM.
The researched two-loop self-energy diagrams include:
1. the two-loop Barr-Zee type diagrams with fermion sub-loop, 2. the two-loop rainbow diagrams with fermion sub-loop and the vector bosons($\gamma$, Z, W).
3. the diamond type diagrams in Ref.\cite{two2,our} possessing large factors.
These diagrams are shown in the Fig.\ref{twolooptu}.
We give the explanation why the particular subset of diagrams is chosen.

1. Fig.\ref{twolooptu}(a), Fig.\ref{twolooptu}(b) and Fig.\ref{twolooptu}(c) are the two-loop Barr-Zee type diagrams,
and their contributions to muon MDM are studied particularly in the work\cite{other2}.
On the supposition $\chi^\pm\sim\chi^0\sim M$, we can obtain very concise results with the factor
$\frac{x_\mu}{x_M^{1/2}x_V^{1/2}}=\frac{m^2_\mu}{Mm_V}$.
$m_V$ represents the mass of heavy vector bosons $m_Z\sim m_W\sim m_V$.
The factor $\frac{m_\mu}{m_V}$ comes from the Higgs lepton vertex $\bar{\mu}-H-\mu$.
This type contributions are considerable.

2. The rainbow diagrams (Fig.\ref{twolooptu}(d) and Fig.\ref{twolooptu}(e)) with heavy fermion sub-loop
 and two vector bosons ZZ(WW, Z$\gamma$, $\gamma\gamma$) can give important corrections to muon MDM\cite{ffa,lepton}.

3. The two-loop self-energy diagrams (Fig.\ref{twolooptu}(f), $\dots$, Fig.\ref{twolooptu}(j)) belong to the diamond type,
where the vector boson couples with external lepton.
This type two-loop diagram studied in this work contains five virtual particles including: one vector boson, two scalars and two fermions.
The corresponding contributions to muon MDM possess two unique factors: $\frac{x_l^{1/2}}{x_M^{1/2}}$ and $\frac{x_l}{x_V}$.
It should be noted that, the total effects from the factor $\frac{x_l^{1/2}}{x_M^{1/2}}$
and the rotation matrixes change into the factor $\frac{x_l}{x_M}\tan\beta$ in the end.
It is similar as the condition of the one-loop results\cite{dabeta2}. These diagrams can also give considerable corrections.

 In fact, there are so many two-loop diagrams that we can not calculate all of them in one work.
 Furthermore, the calculation of the two-loop diagrams is very tedious. So, we study the two-loop SUSY diagrams step by step, and calculate
 several types of two-loop diagrams giving important contributions to muon MDM. Other two-loop diagrams will be researched in our future work.

To obtain the corrections to muon MDM from these two-loop diagrams, we have to resolve the
complicated two-loop integrals. The required steps are the following:


 $\bullet$  We use "momentum expansion" method \cite{zhkrp}, and
assume that all external leptons as well as photon are off-shell,
then expand the amplitude of corresponding triangle diagrams
according to the external momenta of leptons and photon\cite{other2,fengtf04}.
We expand them in
powers of external momenta to the second order. The even rank
tensors in the loop momenta $k_1$, $k_2$ are kept.

 $\bullet$ To simplify the calculation, we use the following formulas of
  the loop momenta $k_1$ and $k_2$\cite{other2,fengtf04}.

 {\small
 \begin{eqnarray}
&&\int\!\!\!\frac{d^Dk_1d^Dk_2}{(2\pi)^{2D}}\frac{(k_{1\mu}k_{1\nu},~k_{1\mu}k_{2\nu})}{\mathcal{D}_0}
\rightarrow \int\!\!\!\frac{d^Dk_1d^Dk_2}{(2\pi)^{2D}} \frac{(k^2,~k_1\cdot k_2)g_{\mu\nu}}{D\mathcal{D}_0},
\nonumber\\
&&\int\!\!\!\frac{d^Dk_1d^Dk_2}{(2\pi)^{2D}}\frac{(k_{1\mu}k_{1\nu}k_{1\rho}k_{1\sigma},~k_{1\mu}k_{1\nu}k_{1\rho}k_{2\sigma})}{\mathcal{D}_0}
\rightarrow\int\!\!\!\frac{d^Dk_1d^Dk_2}{(2\pi)^{2D}}\frac{(k_1^4,~k_1^2k_1\cdot k_2)T_{\mu\nu\rho\sigma}}{D(D+2)\mathcal{D}_0},\nonumber\\
&&\int\!\!\!\frac{d^Dk_1d^Dk_2}{(2\pi)^{2D}}\frac{k_{1\mu}k_{1\nu}k_{2\rho}k_{2\sigma}
}{\mathcal{D}_0}\nonumber\\
 &&\rightarrow\int\frac{d^Dk_1d^Dk_2}{(2\pi)^{2D}}\frac{1}{\mathcal{D}_0}
 \Big(\frac{D(k_1\cdot
k_2)^2-k_1^2k_2^2}{D(D-1)(D+2)}T_{\mu\nu\rho\sigma}-\frac{(k_1\cdot
k_2)^2-k_1^2k_2^2}{D(D-1)}g_{\mu\nu}g_{\rho\sigma}\Big),
\nonumber\\
&&\int\!\!\!\frac{d^Dk_1d^Dk_2}{(2\pi)^{2D}}\frac{k_{1\mu}k_{1\nu}k_{1\rho}k_{1\sigma}
k_{2\alpha}k_{2\beta}}{\mathcal{D}_0}\nonumber\\
 &&\rightarrow\int\!\!\!\frac{d^Dk_1d^Dk_2}{(2\pi)^{2D}}\frac{1}{\mathcal{D}_0}
 \Big(\frac{D(k_1\cdot
k_2)^2k_1^2-k_1^4k_2^2}{D(D+2)(D+4)(D-1)}T_{\alpha\beta\mu\nu\rho\sigma}-\frac{(k_1\cdot
k_2)^2k_1^2-k_1^4k_2^2}{D(D+2)(D-1)}g_{\alpha\beta}T_{\mu\nu\rho\sigma}\Big),\nonumber\\
&&\int\frac{d^Dk_1d^Dk_2}{(2\pi)^{2D}}\frac{k_{1\mu}k_{1\nu}k_{1\rho}k_{2\sigma}
k_{2\alpha}k_{2\beta}}{\mathcal{D}_0}\nonumber\\
 &&\rightarrow\int\!\!\!\frac{d^Dk_1d^Dk_2}{(2\pi)^{2D}}\frac{1}{\mathcal{D}_0}
 \Big(\frac{(D+1)k_1^2k_2^2(k_1\cdot
k_2)-2(k_1\cdot
k_2)^3}{D(D+2)(D+4)(D-1)}T_{\alpha\beta\mu\nu\rho\sigma}-\frac{k_1^2k_2^2(k_1\cdot
k_2)-(k_1\cdot k_2)^3}{D(D+2)(D-1)}\nonumber\\
&&\times[g_{\mu\sigma}(g_{\nu\alpha}
g_{\rho\beta}+g_{\nu\beta}g_{\rho\alpha})+g_{\mu\alpha}(g_{\nu\sigma}g_{\rho\beta}
+g_{\nu\beta}g_{\rho\sigma})+g_{\mu\beta}(g_{\nu\sigma}g_{\rho\alpha}
+g_{\nu\alpha}g_{\rho\sigma})]\Big)\label{tihuan},
 \end{eqnarray}}
where the time-space dimension $D=4-2\epsilon$. The concrete forms of $T_{\mu\nu\rho\sigma}$, $T_{\mu\nu\rho\sigma\alpha\beta}$ and
$\mathcal{D}_0$ are
\begin{eqnarray}
&&T_{\mu\nu\rho\sigma}=g_{\mu\nu}g_{\rho\sigma}+g_{\mu\rho}g_{\nu\sigma}
+g_{\mu\sigma}g_{\nu\rho},\nonumber\\&&
T_{\mu\nu\rho\sigma\alpha\beta}=g_{\mu\nu}T_{\rho\sigma\alpha\beta}+g_{\mu\rho}T_{\nu\sigma\alpha\beta}
+g_{\mu\sigma}T_{\nu\rho\alpha\beta}+g_{\mu\alpha}T_{\nu\rho\sigma\beta}+g_{\mu\beta}T_{\nu\rho\sigma\alpha},\nonumber\\&&
\mathcal{D}_0=(k_1^2-m_1^2)(k_2^2-m_2^2)\Big((k_1-k_2)^2-m_0^2\Big).
\end{eqnarray}

 $\bullet$ Because the integrations are symmetric under the transformation
$k_{1,2}\rightarrow-k_{1,2}$, we abandon the odd rank terms in the
loop momenta. With the decomposition formula
\begin{eqnarray}
&&\frac{1}{(k^2-m_1^2)}\frac{1}{(k^2-m_2^2)}=\frac{1}{m_1^2-m_2^2}\left(
\frac{1}{k^2-m_1^2}-\frac{1}{k^2-m_2^2}\right)\label{chai},\nonumber
\\&&
\frac{1}{(k^2-m^2)^n}=\frac{1}{(n-1)!}\left(\frac{\partial}{\partial
m^2}\right)^{n-1}\frac{1}{k^2-m^2}\label{dao},
\end{eqnarray}
the complicated two-loop integrals can be reduced to the simple form
with one $1/\mathcal{D}_0$.  All the two-loop integrations that we
treat can be simplified to the two-loop vacuum
integrals\cite{npb93tl} and one-loop integrals.

$\bullet$ The two-loop vacuum integral is expressed as
\begin{eqnarray}
&&\Lambda_{{\rm RE}}^{4\epsilon}\int\int{d^Dk_1d^Dk_2\over(2\pi)^{2D}}{1\over
(k_1^2-m_1^2)(k_2^2-m_2^2)((k_1-k_2)^2-m_0^2)}
\nonumber\\&&=
{\Lambda^2\over2(4\pi)^4}{\Gamma^2(1+\epsilon)\over(1-\epsilon)^2}
\Big({4\pi x_{R}}\Big)^{2\epsilon}
\Big\{-{1\over\epsilon^2}\Big(x_0+x_1+x_2\Big)
+{1\over\epsilon}\Big(-(x_0+x_1+x_2)\nonumber\\&&+2(x_0\ln x_0+x_1\ln x_1+x_2\ln x_2)\Big)
+2(x_0\ln x_0+x_1\ln x_1+x_2\ln x_2)
\nonumber\\&&-2(x_0+x_1+x_2)
-x_0\ln^2x_0-x_1\ln^2x_1-x_2\ln^2x_2-\Phi(x_0,x_1,x_2)\Big\},
\label{2l-vacuum}
\end{eqnarray}
with
\begin{eqnarray}
&&\Phi(x,y,z)=(x+y-z)\ln x\ln y+(x-y+z)\ln x\ln z
\nonumber\\
&&\hspace{1.7cm}+(y+z-x)\ln y\ln z+{\rm sign}(\lambda^2)\sqrt{|\lambda^2|}\Psi(x,y,z)\;.
\label{phi}
\end{eqnarray}

The concrete form of $\Psi(x,y,z)$ is:

$1.~~\lambda^2>0,\;\sqrt{y}+\sqrt{z}<\sqrt{x}$:
\begin{eqnarray}
&&\Psi(x,y,z)=2\ln\Big({x+y-z-\lambda\over2x}\Big)
\ln\Big({x-y+z-\lambda\over2x}\Big)-\ln{y\over x}\ln{z\over x}
\nonumber\\
&&\hspace{2.2cm}
-2L_{i_2}\Big({x+y-z-\lambda\over2x}\Big)
-2L_{i_2}\Big({x-y+z-\lambda\over2x}\Big)+{\pi^2\over3}\;,
\label{aeq2}
\end{eqnarray}
with $L_{i_2}(x)$ representing the spence function;

 $2.~~\lambda^2>0,\;\sqrt{x}+\sqrt{z}<\sqrt{y}$:
\begin{eqnarray}
&&\Psi(x,y,z)={\rm Eq.}(\ref{aeq2})(x\leftrightarrow y)\;;
\label{aeq3}
\end{eqnarray}

$3.~~\lambda^2>0,\;\sqrt{x}+\sqrt{y}<\sqrt{z}$:
\begin{eqnarray}
&&\Psi(x,y,z)={\rm Eq.}(\ref{aeq2})(x\leftrightarrow z)\;;
\label{aeq4}
\end{eqnarray}

 $4.~~\lambda^2<0$:
\begin{eqnarray}
&&\Psi(x,y,z)=2\Big\{Cl_2\Big(2\arccos(
{-x+y+z\over2\sqrt{yz}})\Big)
+Cl_2\Big(2\arccos({x-y+z\over2\sqrt{xz}})\Big)
\nonumber\\
&&\hspace{2.2cm}
+Cl_2\Big(2\arccos({x+y-z\over2\sqrt{xy}})\Big)\Big\}\;,
\label{aeq12}
\end{eqnarray}
with $Cl_2(x)$ denoting the Clausen function.

With the supposition $z\ll x\sim y$, the function $\Phi(x,y,z)$ is much simplified as
\begin{eqnarray}
&&\Phi(x,y,z)\simeq 2 x \ln ^2x+z
   \Big(2 (\ln z-2)\ln x -\ln^2x+4 \ln z\Big)\nonumber\\&&\hspace{1.8cm}
   +\frac{z^2 (3 \ln x-3 \ln z+2)}{9 x}+\frac{z^3 (15 \ln x-15 \ln z+1)}{450 x^2}+\dots.
\end{eqnarray}

\begin{figure}[h]
\vspace{-1.5cm}
\setlength{\unitlength}{1mm}
\centering
\includegraphics[width=7.0in]{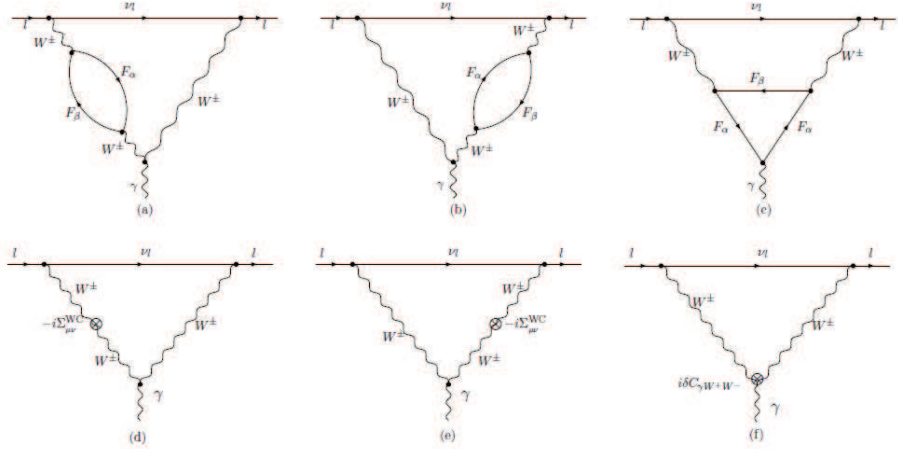}
\vspace{-17cm}
\caption[]{The two-loop triangle diagrams with a closed heavy
fermion sub-loop correspond to the two-loop self-energy diagram Fig.\ref{twolooptu}(d).
The diagrams (d, e, f) give counter terms to cancel the UV-divergences produced from the UV-divergent sub-diagrams in (a, b, c) respectively.}\label{wwst}
\end{figure}

Here, we give an example for the calculation of the contributions to $a_\mu$ from the two-loop self-energy diagram. For Fig.\ref{twolooptu}(d),
  a closed heavy fermion loop is inserted into the self-energy of W gauge boson.
 The heavy fermions in the sub-loop are chargino and neutralino.
 In the Fig.\ref{wwst}, the diagrams Fig.\ref{wwst}(a), Fig.\ref{wwst}(b),
 Fig.\ref{wwst}(c) are the two-loop triangle diagrams and they have UV-divergent
 terms which are caused by the UV-divergent sub-diagrams. Their counter terms are denoted by
 the diagrams Fig.\ref{wwst}(d), Fig.\ref{wwst}(e) and Fig.\ref{wwst}(f) respectively. The UV-divergent
 term of Fig.\ref{wwst}(a) comes from the W one-loop self-energy diagram with heavy virtual fermion. Fig.\ref{wwst}(d) is the counter term of Fig.\ref{wwst}(a),
 whose condition is same as that of Fig.\ref{wwst}(e) and Fig.\ref{wwst}(b). In Fig.\ref{wwst}(c), $F_\alpha$ denotes chargino and
 $F_\beta$ represents neutralino. Its sub-loop is the one-loop
 diagram of the vertex $\gamma W^+W^-$ producing UV-divergent term. Fig.\ref{wwst}(f) represents the counter term of Fig.\ref{wwst}(c).

 The sum of the Feynman amplitudes for diagrams in Fig.\ref{wwst}
satisfies the Ward-identity, which is required by the QED gauge invariance
\begin{eqnarray}
&&q_\mu\Gamma^\mu_{{ WW}}(p,q)=e[\Sigma_{WW}(p+q)
-\Sigma_{WW}(p)]\;,
\end{eqnarray}
with $\Gamma_{WW}^\mu$ denoting the sum of amplitudes for the diagrams Fig.\ref{wwst}(a), Fig.\ref{wwst}(b), Fig.\ref{wwst}(c).
Correspondingly, $\Sigma_{WW}$ denotes the amplitude of W self-energy
diagram. In general way, the unrenormalized W self-energy can be expressed as
\begin{eqnarray}
&&\Sigma_{\mu\nu}^{WW}(p,\Lambda_{{RE}})=\Lambda^2\mathcal{A}_0^{W}g_{\mu\nu}+\Big(\mathcal{A}_1^{W}
+{p^2\over\Lambda^2}\mathcal{A}_2^{W}+\cdots\Big)(p^2g_{\mu\nu}-p_\mu p_\nu)
\nonumber\\
&&\hspace{2.5cm}
+\Big(\mathcal{B}_1^{W}+{p^2\over\Lambda^2}\mathcal{B}_2^{W}+\cdots\Big)p_\mu p_\nu.
\label{eq-w1}
\end{eqnarray}
The form factors $\mathcal{A}_{0,1,2}^{W}$ and $\mathcal{B}_{1,2}^{W}$ are obtained after performing loop integration, and
they are function of the virtual particle masses and renormalization scale.

 The counter terms of W self-energy are shown in the following form
\begin{eqnarray}
&&\Sigma_{{\mu\nu}}^{W,C}(p,\Lambda_{{RE}})=-\Big[\delta m_{W}^2(\Lambda_{{RE}})
+m_{W}^2\delta Z_{W}(\Lambda_{{ RE}})\Big]g_{\mu\nu}
-\delta Z_{W}(\Lambda_{{RE}})\Big[p^2g_{\mu\nu}-p_\mu p_\nu\Big].
\label{eq-w2}
\end{eqnarray}
From $\Sigma_{\mu\nu}^{WW}(p,\Lambda_{{RE}})$ and $\Sigma_{{\mu\nu}}^{W,C}(p,\Lambda_{{RE}})$,
one can obtain the renormalized self-energy
\begin{eqnarray}
&&\hat{\Sigma}_{{\mu\nu}}^{W}(p,\Lambda_{{RE}})=
\Sigma_{{\mu\nu}}^W(p,\Lambda_{{ RE}})
+\Sigma_{{\mu\nu}}^{W,C}(p,\Lambda_{{RE}}).
\end{eqnarray}

The condition for the on-shell external gauge boson W reads as
\begin{eqnarray}
&&\hat{\Sigma}_{{\mu\nu}}^{W}(p,m_{{W}})\epsilon^\nu(p)\Big|_{p^2=m_{W}^2}=0
\;,\nonumber\\
&&\lim\limits_{p^2\rightarrow m_{W}^2}{1\over p^2-m_{W}^2}
\hat{\Sigma}_{{\mu\nu}}^{W}(p,m_{W})\epsilon^\nu(p)=\epsilon_{\mu}(p)\;,
\label{eq-w4}
\end{eqnarray}
with $\epsilon(p)$ denoting the polarization vector of W gauge boson.
From Eq.(\ref{eq-w1}), Eq.(\ref{eq-w2}) and Eq.(\ref{eq-w4}),
 the counter terms for the W self-energy are deduced in on-shell scheme
\begin{eqnarray}
&&\delta Z_{W}(m_{W})=\mathcal{A}_1^{W}+x_{W}\mathcal{A}_2^{W},~~~
\delta m_{W}^{2}(m_{W})=\mathcal{A}_0^{W}\Lambda^2
-m_{W}^2\delta Z_{W}.
\label{eq-w5}
\end{eqnarray}

There is the $\gamma W^+W^-$ vertex at tree level, whose counter term is derived in the following form
\begin{eqnarray}
&&i\delta C_{\gamma W^+W^-}=ie\cdot\delta Z_{W}(\Lambda_{RE})
\Big[g_{\mu\nu}(k_1-k_2)_\rho+g_{\nu\rho}(k_2-k_3)_\mu+g_{\rho\mu}(k_3-k_1)_\nu\Big].
\label{eq-w6}
\end{eqnarray}
Here, $k_1$ and $k_2$ denote the incoming momenta of $W^\pm$. While $k_3$ denotes the incoming momentum of photon.
$\mu,\;\nu,\;\rho$ are the corresponding Lorentz indices.

After tedious calculation, we obtain the analytic results
of the diagrams in the Fig.\ref{wwst}, whose sum is finite and very complex. Because the full analytic
 results take up a lot of space, we do not show them here.
 In order to make the analytic results more concise and practice, the full analytic
 results are expended in the condition $m_{F_\alpha}\sim m_{F_\beta}\gg m_W\gg m_\mu$.
In the end, the very complex results are much simplified and they are shown as $a^{2L,WW}_\mu$ in the following parts noted by Eq.(\ref{tlww}).
In the similar way, the other two-loop diagrams are also deduced and expanded. In the end, the simplified analytic results are
obtained and shown in this work.

With the assumption $m_{F_1}=m_{F_2}\gg m_W$, the results \cite{ffa} for the Fig.\ref{twolooptu} (a) can be simplified as
\begin{eqnarray}
&&a_\mu^{2L,~WH}=\frac{eH_{\bar{\mu}H\nu}^L}{512\sqrt{2}\pi^4s_W}\sum_{F_1=\chi^{\pm}}\sum_{F_2=\chi^0}\frac{x_\mu^{1/2}}{ x^{1/2}_{F_1}}\Big\{\frac{199}{36}\Re(H_{H\bar{F}_1F_2}^LH_{W\bar{F}_2F_1}^L+H_{H\bar{F}_1F_2}^RH_{W\bar{F}_2F_1}^R)
\nonumber\\&&\hspace{1.6cm}+\Big[\frac{13}{3}+2
(\ln{x_{F_1}}-\varrho_{1,1}(x_W,x_{H^\pm}))\Big]\Re(H_{H\bar{F}_1F_2}^LH_{W\bar{F}_2F_1}^R+H_{H\bar{F}_1F_2}^RH_{W\bar{F}_2F_1}^L)
\nonumber\\
&&\hspace{1.6cm}+\Big[\frac{4}{3}(\ln{x_{F_1}}-\varrho_{1,1}(x_W,x_{H^\pm}))-\frac{16}{9}\Big]\Re(H_{H\bar{F}_1F_2}^LH_{W\bar{F}_2F_1}^L\hspace{-0.1cm}
-\hspace{-0.1cm}H_{H\bar{F}_1F_2}^RH_{W\bar{F}_2F_1}^R)
\nonumber\\
&&\hspace{1.6cm}+\Big[\frac{2}{9}-\frac{8}{3}(\ln{x_{F_1}}-\varrho_{1,1}(x_W,x_{H^\pm}))\Big]\Re(H_{H\bar{F}_1F_2}^LH_{W\bar{F}_2F_1}^R\hspace{-0.1cm}
-\hspace{-0.1cm}H_{H\bar{F}_1F_2}^RH_{W\bar{F}_2F_1}^L)\Big\},
\\
&&a_\mu^{2L,~WG}
=\frac{eH_{\bar{\mu}G\nu}^L}{512\sqrt{2}\pi^4s_W}\sum_{F_1=\chi^{\pm}}\sum_{F_2=\chi^0}\frac{x_\mu^{1/2}}{ x^{1/2}_{F_1}}\Big\{\frac{199}{36}\Re(H_{G\bar{F}_1F_2}^LH_{W\bar{F}_2F_1}^L+H_{G\bar{F}_1F_2}^RH_{W\bar{F}_2F_1}^R)
\nonumber\\&&\hspace{1.6cm}+\Big[\frac{7}{3}+2
(\ln{x_{F_1}}-\ln x_W)\Big]\Re(H_{G\bar{F}_1F_2}^LH_{W\bar{F}_2F_1}^R+H_{G\bar{F}_1F_2}^RH_{W\bar{F}_2F_1}^L)
\nonumber\\
&&\hspace{1.6cm}+\Big[\frac{4}{3}(\ln{x_{F_1}}-\ln x_W)-\frac{28}{9}\Big]\Re(H_{G\bar{F}_1F_2}^LH_{W\bar{F}_2F_1}^L\hspace{-0.1cm}
-\hspace{-0.1cm}H_{G\bar{F}_1F_2}^RH_{W\bar{F}_2F_1}^R)
\nonumber\\
&&\hspace{1.6cm}+\Big[\frac{26}{9}-\frac{8}{3}(\ln{x_{F_1}}-\ln x_W)\Big]\Re(H_{G\bar{F}_1F_2}^LH_{W\bar{F}_2F_1}^R\hspace{-0.1cm}
-\hspace{-0.1cm}H_{G\bar{F}_1F_2}^RH_{W\bar{F}_2F_1}^L)\Big\}.\label{awg}
\end{eqnarray}

with $\varrho_{1,1}(x,y)=\frac{x\ln x-y\ln y}{x-y}$. The Feynman rules of F-H-F and F-W-F vertexes are written in the following form,
\begin{eqnarray}
&&\mathcal{L}_{H(G)\bar{F}_1F_2}=i\bar{F_1}(H_{H\bar{F}_1F_2}^{L}P_L+H_{H\bar{F}_1F_2}^{R}P_R)F_2H^{\pm}
\nonumber\\&&\hspace{2.0cm}+i\bar{F_1}(H_{G\bar{F}_1F_2}^{L}P_L+H_{G\bar{F}_1F_2}^{R}P_R)F_2G^{\pm},\nonumber\\&&
 \mathcal{L}_{W\bar{F}_1F_2}=i\bar{F_1}(H_{W\bar{F}_1F_2}^{L}\gamma_\mu P_L+H_{W\bar{F}_1F_2}^{R}\gamma_\mu P_R)F_2W^{\pm\mu}.
\end{eqnarray}
One can find the concrete forms of $H_{H(G)\bar{F}_1F_2}^{L,R}$ and $H_{W\bar{F}_1F_2}^{L,R}$ in the Ref.\cite{slh}.

Using similar assumption $m_{F_1}=m_{F_2}\gg m_{h^0}$, one can simplify the two-loop Barr-Zee type
diagrams contributing to the muon MDM for the
Figs. \ref{twolooptu} (b) and \ref{twolooptu} (c)
\begin{eqnarray}
&&a_\mu^{2L,~\gamma h^0}=\frac{e^2}{64\sqrt{2}\pi^4}H_{h^0\bar{\mu}\mu}\sum_{F_1=F_2=\chi^\pm}\frac{x_\mu^{1/2}}{ x^{1/2}_{F_1}}
\Re(H_{h^0\bar{F}_1F_2}^L)\Big[1+\ln\frac{x_{F_1}}{x_{h^0}}\Big],
\\
&&a_\mu^{2L,~Zh^0}=\frac{\sqrt{2}}{512\pi^4}\sum_{F_1=F_2=\chi^{\pm},\chi^0}
H_{h^0\bar{\mu}\mu}\frac{x_\mu^{1/2}}{ x^{1/2}_{F_1}}\Big[\varrho_{1,1}(x_Z,x_{h^0})-\ln{x_{F_1}}-1\Big]
\nonumber\\&&\hspace{1.6cm}\times(H^L_{Z\bar{\mu}\mu}+H^R_{Z\bar{\mu}\mu})\Re(H_{h^0\bar{F}_1F_2}^LH_{Z\bar{F}_2F_1}^L+H_{h^0\bar{F}_1F_2}^RH_{Z\bar{F}_2F_1}^R).
\end{eqnarray}


The couplings of the CP-odd neutral bosons(the neutral Goldstone boson) with fermions  are written as
\begin{eqnarray}
&&\mathcal{L}_{A^0(G^0)\mu\mu}=\bar{l}H^{\gamma_5}_{A^0\bar{\mu}\mu}\gamma_5lA^0+\bar{l}H^{\gamma_5}_{G^0\bar{\mu}\mu}\gamma_5lG^0,\nonumber\\&&
 \mathcal{L}_{A^0(G^0)F_1F_2}=\bar{F}_1(H_{A^0\bar{F}_1F_2}^{L}P_L+H_{A^0\bar{F}_1F_2}^{R}P_R)F_2A^0\nonumber\\&&\hspace{2.2cm}+
 \bar{F}_1(H_{G^0\bar{F}_1F_2}^{L}P_L+H_{G^0\bar{F}_1F_2}^{R}P_R)F_2G^0+h.c.
\end{eqnarray}

In the same way, the contributions of Fig. 2b and Fig. 2c with $A^0$ and $G^0$ instead of $h^0$ are obtained
\begin{eqnarray}
&&a_\mu^{2L,~\gamma A^0}=-\frac{e^2Q_{F_1}^2}{64\sqrt{2}\pi^4}
\sum_{F_1=\chi^\pm}\frac{x_\mu^{1/2}}{ x^{1/2}_{F_1}}
\Re(H^{\gamma_5}_{A^0\bar{\mu}\mu}H_{A^0\bar{F}_1F_1}^L)(1+\ln\frac{x_{F_1}}{x_{A^0}}),\label{aga}
\\&&
a_\mu^{2L,~\gamma G^0}=-\frac{e^2Q_{F_1}^2}{64\sqrt{2}\pi^4}
\sum_{F_1=\chi^\pm}\frac{x_\mu^{1/2}}{ x^{1/2}_{F_1}}
\Re(H^{\gamma_5}_{G^0\bar{\mu}\mu}H_{G^0\bar{F}_1F_1}^L)(1+\ln\frac{x_{F_1}}{x_{Z}}),\label{agg}
\\
&&a_\mu^{2L,~ZA^0}=-\frac{\sqrt{2}}{256\pi^4}\sum_{F_1=F_2=\chi^{\pm},\chi^0}
\frac{x_\mu^{1/2}}{ x^{1/2}_{F_1}}(H^L_{Z\bar{\mu}\mu}+H^R_{Z\bar{\mu}\mu})
\Re \Big((H_{A^0\bar{F}_1F_2}^LH_{Z\bar{F}_2F_1}^L\nonumber\\&&\hspace{1.6cm}
+H_{A^0\bar{F}_1F_2}^RH_{Z\bar{F}_2F_1}^R)H^{\gamma_5}_{A^0\bar{\mu}\mu}\Big)[\varrho_{1,1}(x_Z,x_{A^0})-\ln{x_{F_1}}-1],\label{aza}
\\&&
a_\mu^{2L,~ZG^0}=-\frac{\sqrt{2}}{256\pi^4}\sum_{F_1=F_2=\chi^{\pm},\chi^0}
\frac{x_\mu^{1/2}}{ x^{1/2}_{F_1}}(H^L_{Z\bar{\mu}\mu}+H^R_{Z\bar{\mu}\mu})
\nonumber\\&&\hspace{1.6cm}\times
\Re [H^{\gamma_5}_{G^0\bar{\mu}\mu}(H_{G^0\bar{F}_1F_2}^LH_{Z\bar{F}_2F_1}^L
+H_{G^0\bar{F}_1F_2}^RH_{Z\bar{F}_2F_1}^R)]\ln \frac{x_Z}{x_{F_1}}.\label{azg}
\end{eqnarray}
They have a suppression factor$\frac{m_\mu}{m_W}\sim\frac{x_\mu^{1/2}}{ x^{1/2}_{V}}$ from
the vertex couplings $H_{\bar{\mu}G\nu}^L$, $H^{\gamma_5}_{A^0\bar{\mu}\mu}$, $H^{\gamma_5}_{G^0\bar{\mu}\mu}$ respectively.

For the two-loop rainbow diagrams with two vector bosons $(\gamma,\gamma)$ and $(\gamma, Z)$,  their contributions are simplified with the
 assumption $m_{F_1}=m_{F_2}\gg m_W\sim m_Z$, and the simplified results  are suppressed by the small factor $\frac{x_{\mu}}{x_{F_1}}$, as discussed in Ref. \cite{our}.
 Therefore, we can neglect their corrections safely. With the same assumption, the two-loop rainbow diagrams with two vector bosons (W, W) are deduced here
  \begin{eqnarray}
&&a_\mu^{2L,~WW}=\frac{e^2}{1536\pi^4s_W^2}\frac{x_\mu}{x_W}\sum_{F_1=\chi^{\pm}}\sum_{F_2=\chi^0}
\Big\{-31(|H_{W\bar{F}_1F_2}^L|^2+|H_{W\bar{F}_1F_2}^R|^2)
\nonumber\\
&&-12(|H_{W\bar{F}_1F_2}^L|^2-|H_{W\bar{F}_1F_2}^R|^2)+11\Re(H_{W\bar{F}_1F_2}^{R*}H_{W\bar{F}_1F_2}^L)\Big\}.\label{tlww}
\end{eqnarray}

In the similar way, the corrections from rainbow diagrams with two Z vector bosons are also
simplified and only the terms with the largest factor $\frac{x_\mu}{x_Z}$ are kept.
\begin{eqnarray}
&&a_{\mu}^{2L,~ZZ}=\frac{1}{1024\pi^4} \frac{x_\mu}{x_Z}\sum_{F_1=F_2=\chi^\pm}\Big\{
-6\Big(|H^L_{Z\bar{F}_1F_2}|^2+|H^R_{Z\bar{F}_1F_2}|^2\Big)\Big(|H^L_{Z\bar{\mu}\mu}|^2
+|H^R_{Z\bar{\mu}\mu}|^2\Big)\nonumber\\&&\times(2\ln x_{F_1}+5)
+16\Big(|H^L_{Z\bar{F}_1F_2}|^2+|H^R_{Z\bar{F}_1F_2}|^2\Big)H^L_{Z\bar{\mu}\mu}H^R_{Z\bar{\mu}\mu}
[(\ln x_{F_1}+2)\ln \frac{x_{F_1}}{ x_Z} +2]\Big\}.
\end{eqnarray}
The diagrams of the form Fig.
2e with $Z\gamma~ (\gamma\gamma)$ exchange instead
of $ZZ $ exchange are calculated\cite{slh,ffa} and
the simplified results read as
\begin{eqnarray}
&&a_\mu^{2L,\gamma Z}=\frac{Q_{F_1}m_\mu^2e^2}{256\pi^4}(H^R_{Z\bar{\mu}\mu}-H^L_{Z\bar{\mu}\mu})\sum_{F_1=F_2=\chi^\pm}
\frac{\Re(H_{Z\bar{F}_1F_2}^L-H_{Z\bar{F}_1F_2}^R)}{m_{F_1}^2}
\Big[35+\ln\frac{x_{F_1}}{x_Z}\Big],\\&&
a_\mu^{2L,\gamma\gamma}=\frac{e^4Q_{F_1}^2}{720\pi^4\sin^2\theta_W}\sum_{F_1=F_2=\chi^\pm}\frac{m_\mu^2}{m_{F_1}^2}.
\end{eqnarray}

The Figs.\ref{twolooptu}(f, g, h, i, j) have their H.C. diagrams, which give same contributions to muon MDM. So, we do not plot them here.
 After tedious calculation and simplification, we obtain the
analytic results in the following form. For the Fig. \ref{twolooptu} (f), we keep the terms as
\begin{eqnarray}
&&a^{2L,\;Z\tilde{\nu}\chi^{\pm}}_{\mu}=\frac{1}{768\pi^4}
\sum_{i=1}^2\sum_{j,k=1}^6 G_{Z\tilde{\nu}^I_j\tilde{\nu}^R_k}\bigg\{\frac{x_{\mu}}{x_Z}\Big[-4\Re\Big(H^L_{Z\bar{\mu}\mu}
H^{L}_{\mu\bar{\chi}_i^\pm\tilde{\nu}^I_j}H^{L*}_{\mu\bar{\chi}^\pm_i\tilde{\nu}^R_k}+H^R_{Z\bar{\mu}\mu}
H^{R}_{\mu\bar{\chi}_i^\pm\tilde{\nu}^I_j}H^{R*}_{\mu\bar{\chi}^\pm_i\tilde{\nu}^R_k}\Big)\nonumber\\&&-\Re\Big(H^L_{Z\bar{\mu}\mu}
H^{R}_{\mu\bar{\chi}_i^\pm\tilde{\nu}^I_j}H^{R*}_{\mu\bar{\chi}^\pm_i\tilde{\nu}^R_k}+H^R_{Z\bar{\mu}\mu}
H^{L}_{\mu\bar{\chi}_i^\pm\tilde{\nu}^I_j}H^{L*}_{\mu\bar{\chi}^\pm_i\tilde{\nu}^R_k}\Big)(6 \ln x_{\mu}-10)\Big]
\nonumber\\&&
+(H^L_{Z\bar{\mu}\mu}+H^R_{Z\bar{\mu}\mu})\Re\Big(H^{R}_{\mu\bar{\chi}^\pm_i\tilde{\nu}^I_j}H^{L*}_{\mu\bar{\chi}^\pm_i\tilde{\nu}^R_k}\Big)
\frac{x_{\mu}^{1/2}}{x_{\chi^{\pm}_i}^{1/2}}\Big(\ln x_{\chi^\pm_i}-2 \ln x_Z-\frac{35}{12}\Big)\bigg\}+(\tilde{\nu}^I\leftrightarrow\tilde{\nu}^R)
.
\end{eqnarray}
$G_{VS^*_1S_2}$ is the coupling constant for one vector boson and two scalars with the general form
\begin{eqnarray}
\mathcal{L}_{{VS^*_1S_2}}=S_1^*G_{VS^*_1S_2}(-p_{\mu}^{S_1^*}+p_\mu^{S_2})S_2V^\mu.
\end{eqnarray}

For the Fig.\ref{twolooptu}(h) with vector boson Z and the Fig.\ref{twolooptu}(i) with gauge boson $\gamma$,  the results are deduced
\begin{eqnarray}
&&a^{2L,\;Z\tilde{L}\chi^{0}}_{\mu}=\frac{-1}{1536 \pi ^4}\sum_{s,t=1}^6\sum_{j=1}^8\limits
\bigg\{4\Re\Big(H^{R*}_{\mu\bar{\chi}^0_j\tilde{L}_t}{G}_{Z\tilde{L}_t\tilde{L}_s}
H^{R}_{\mu\bar{\chi}^0_j\tilde{L}_s}\Big)\frac{x_{\mu}}{x_Z}\Big[4H^R_{Z\bar{\mu}\mu}+H^L_{Z\bar{\mu}\mu}(3\ln x_{\mu}-5)\Big]
\nonumber\\&&\hspace{2.2cm}
+4\Re\Big(H^{L*}_{\mu\bar{\chi}^0_j\tilde{L}_t}{G}_{Z\tilde{L}_t\tilde{L}_s}
H^{L}_{\mu\bar{\chi}^0_j\tilde{L}_s}\Big)\frac{x_{\mu}}{x_Z}
\Big[4 H^L_{Z\bar{\mu}\mu} +H^R_{Z\bar{\mu}\mu}( 3 \ln x_{\mu}-5)\Big]\nonumber\\&&\hspace{2.2cm}
-\Re\Big(H^{R*}_{\mu\bar{\chi}^0_j\tilde{L}_t}{G}_{Z\tilde{L}_t\tilde{L}_s}
H^{L}_{\mu\bar{\chi}^0_j\tilde{L}_s}+
H^{L*}_{\mu\bar{\chi}^0_j\tilde{L}_t}{G}_{Z\tilde{L}_t\tilde{L}_s}
H^{R}_{\mu\bar{\chi}^0_j\tilde{L}_s}\Big)\nonumber\\&&\hspace{2.2cm}\times(H^L_{Z\bar{\mu}\mu}+H^R_{Z\bar{\mu}\mu})
\frac{x_{\mu}^{1/2}}{x_{\chi^{0}_j}^{1/2}}\Big(8\ln x_{\chi^{0}_j}-10 \ln x_Z+\frac{287}{12}
  \Big)\bigg\},\\
&&a^{2L,\;\gamma\tilde{L}\chi^{0}}_{\mu}=\frac{e^2}{384\pi^4}\sum_{s=1}^6\sum_{j=1}^8\limits
\Re\Big(H^{R*}_{\mu\bar{\chi}^0_j\tilde{L}_s}H^{L}_{\mu\bar{\chi}^0_j\tilde{L}_s}\Big)
\frac{x_{\mu}^{1/2}}{x_{\chi^{0}_j}^{1/2}}
\Big[10 \log x_{\mu}-8 \log x_{\chi^{0}_j}-\frac{289}{12}\Big].
\end{eqnarray}

 Both Fig.\ref{twolooptu}(g) and Fig.\ref{twolooptu}(j) have W boson, whose contributions to muon MDM are shown here
\begin{eqnarray}
&& a^{2L,\;W\tilde{L}\tilde{\nu}\chi^{0}}_{\mu}=\frac{1}{768\pi ^4}\sum_{j=1}^8\sum_{i,k=1}^6 H^L_{W\bar{\nu} \mu}
\bigg\{-15\frac{x_{\mu}}{x_W}\Re\Big(H^{L*}_{\mu\bar{\chi}^0_j\tilde{L}_k}H^{L}_{\nu\bar{\chi}^0_j\tilde{\nu}^I_i}G_{W\tilde{L}_k\tilde{\nu}^I_i}
\Big)\nonumber\\&&+\Re\Big(H^{R*}_{\mu\bar{\chi}^0_j\tilde{L}_k}H^{L}_{\nu\bar{\chi}^0_j\tilde{\nu}_i^I}G_{W\tilde{L}_k\tilde{\nu}^I_i}
\Big)\frac{x_{\mu}^{1/2}}{x_{\chi^{0}_j}^{1/2}}\Big(\frac{29}{12} -2\ln x_W+4\ln x_{\chi_j^0}\Big)
 \bigg\}+(\tilde{\nu}^I\leftrightarrow \tilde{\nu}^R),
\nonumber\\
&& a^{2L,\;W\tilde{L}\tilde{\nu}\chi^{-}}_{\mu}=-\frac{1}{1536\pi^4}\sum_{j=1}^2\sum_{i,k=1}^6H^L_{W\bar{\nu} \mu}
\bigg\{-18 \frac{x_{\mu}}{x_W}\Re\Big(H^{L*}_{\mu\bar{\chi}^\pm_j\tilde{\nu}^I_i}H^{L}_{\nu\bar{\chi}^\pm_j\tilde{L}_k}G_{W\tilde{L}_k\tilde{\nu}^I_i}
\Big)\nonumber\\&&+\Re\Big(H^{R*}_{\mu\bar{\chi}^\pm_j\tilde{\nu}^I_i}H^{L}_{\nu\bar{\chi}^\pm_j\tilde{L}_k}G_{W\tilde{L}_k\tilde{\nu}^I_i}
\Big)\frac{x_{\mu}^{1/2}}{x_{\chi^{\pm}_j}^{1/2}}\Big(2 \ln x_{\chi_j^\pm}-4 \ln x_W-\frac{7}{6}\Big)\bigg\}+(\tilde{\nu}^I\leftrightarrow \tilde{\nu}^R).
\label{result-zms}
\end{eqnarray}
The corrections to muon MDM from the researched two-loop diagrams are
\begin{eqnarray}
&& a_\mu^{2L}=a_\mu^{2L,~WH}+a_\mu^{2L,~WG}+a_\mu^{2L,~\gamma h_0}+a_\mu^{2L,~\gamma G_0}
+a_\mu^{2L,~\gamma A_0}+a_\mu^{2L,~Z h_0}\nonumber\\&&\hspace{1.0cm}+a_\mu^{2L,~Z G_0}+a_\mu^{2L,~Z A_0}+a_{\mu}^{2L,~WW}
+a_{\mu}^{2L,~ZZ}+a_{\mu}^{2L,~Z\gamma}+a_{\mu}^{2L,~\gamma\gamma}\nonumber\\&&\hspace{1.0cm}
+a^{2L,Z\tilde{\nu}\chi^{\pm}}_{\mu}+a^{2L,Z\tilde{L}\chi^{0}}_{\mu}+a^{2L,\gamma\tilde{L}\chi^{0}}_{\mu}
+a^{2L,W\tilde{L}\tilde{\nu}\chi^{0}}_{\mu}+a^{2L,W\tilde{L}\tilde{\nu}\chi^{-}}_{\mu}.
\end{eqnarray}
At two-loop level, the total results are the sum of one-loop results and two-loop results
\begin{eqnarray}
&& a_\mu=a_{\mu}^{1L}+a_{\mu}^{2L}.
\end{eqnarray}

\section{numerical results}
  In the $U(1)_X$SSM, we have researched several processes\cite{ZSMJHEP,slh}. In this section of the numerical results,
  we  consider the experimental constraints from the lightest CP-even Higgs $h^0$ mass ($\thickapprox$125 GeV) and $h^0$ decays including $h^0\rightarrow \gamma+\gamma$,  $h^0\rightarrow Z+Z$ and $h^0\rightarrow W+W$\cite{2020pdg}.
    The mass constraint for the $Z^\prime$ boson from LHC experiments is more severe than the limits from the low energy data.
  To satisfy the $M_{Z^\prime}$ constraint, we take the parameters to obtain $M_{Z^{\prime}}> 5.1$ TeV\cite{Zp5d1}.
At 99\% CL\cite{UPbmzgx}, the ratio ($M_{Z^\prime}/g_X$) between $M_{Z^\prime}$ and its gauge
coupling should be not smaller than 6 TeV.  To satisfy the LHC experimental data, we take $\tan \beta_\eta< 1.5$ \cite{TBnew}.
The papers\cite{wx1,wx2,wx3,wx4,wx5,wx6,wx7} perform a detailed recasting of LHC limits.
We take into account the experimental constraints on masses of the new
particles to make the scalar lepton masses larger than 700 GeV.
and chargino masses larger than 1100 GeV.

 Considering the above experimental constraints, we adopt the following parameters in the numerical calculation.
\begin{eqnarray}
&&\lambda_H=\kappa=0.1,~\lambda_C=-0.1, ~v_S=3.6~{\rm TeV},  ~\mu=0.9~{\rm TeV},~M_S=1~{\rm TeV},\nonumber\\&&
\xi=17~{\rm TeV},~\tan{\beta_\eta}=1.05, ~
l_W=B_\mu=B_S=0.1~{\rm TeV^2},~T_{\lambda_C}=150~{\rm GeV}, \nonumber\\&&T_{X11}=T_{X22}=T_{X33}=10~{\rm GeV},~m_{S}^2=0.2~{\rm TeV^2},  ~T_{\lambda_H}=T_\kappa=1~{\rm TeV},
\nonumber\\&&Y_{X11}=Y_{X22}=Y_{X33}=0.04,
~T_{E11}=T_{E22}=T_{E33}=0.1~{\rm TeV}, ~M_2=1.2~{\rm TeV}.
\end{eqnarray}
To simplify the numerical discussion, we use the following relations
\begin{eqnarray}
&&M_{Eii}=M_E, ~M_{Lii}=M_L, ~M_{\nu ii}=M_{NU},~T_{\nu ii}=T_{nu}, ~(i=1,2,3).
\end{eqnarray}

\subsection{The numerical results with one or two variables}

\begin{figure}[t]
\begin{center}
\begin{minipage}[c]{0.48\textwidth}
\includegraphics[width=2.9in]{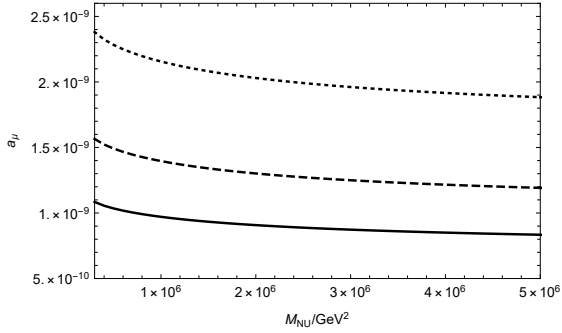}
\end{minipage}%
\caption{$a_\mu$ versus $M_{NU}$. The solid (dashed, dotted) line corresponds to the results with $\tan{\beta}=20~(30, ~40)$.} \label{Mnu}
\end{center}
\end{figure}

In this subsection, we use the parameters as
$M_1= 4 ~{\rm TeV}, M_{BL} = 1~{\rm TeV},~ g_X=0.3,~ g_{YX} = 0.07,~M_E = 0.3 ~{\rm TeV}^2,~T_{nu}= 3~{\rm TeV}$.
$M_{NU}$ are the parameters in the diagonal elements of scalar neutrino(CP-even and CP-odd) mass squared matrix.
Therefore, $M_{NU}$ can affect the masses of scalar neutrino strongly.
With the parameters $M_{BB^\prime} = 3 ~{\rm TeV}$ and $M_L= 0.6 ~{\rm TeV}^2$, we plot $a_\mu$ versus $M_{NU}$ in the Fig.\ref{Mnu},
where the solid (dashed, dotted) line corresponds to the results as $\tan{\beta}=20~(30, ~40)$. The three lines are all
decreasing functions as $M_{NU}$ turn large. That is to say heavy scalar neutrino suppresses the
SUSY contributions to muon g-2. This characteristic
is similar as the MSSM condition, and it is generic for the SUSY models, because the loop corrections from SUSY particles
have the factor $\frac{m^2_\mu}{M_{SUSY}^2}$. The dotted line($\tan\beta=40$) is upon the dashed line($\tan\beta=30$),
and the dashed line is upon the solid line($\tan\beta=20$).
It implies that $\tan\beta$ is a sensitive parameter and larger $\tan\beta$ leads to larger $a_\mu$.
It is consistent with the one-loop results obtained by MIA. The simplified one-loop
results shown as Eq.(\ref{amuS}) are proportional to $\tan\beta$, which is similar as the MSSM condition. The biggest value of the
dotted line can reach $2.4\times10^{-9}$, that can well
compensate the departure between experiment value and the theoretical prediction of SM.

\begin{figure}[t]
\begin{center}
\begin{minipage}[c]{0.48\textwidth}
\includegraphics[width=2.9in]{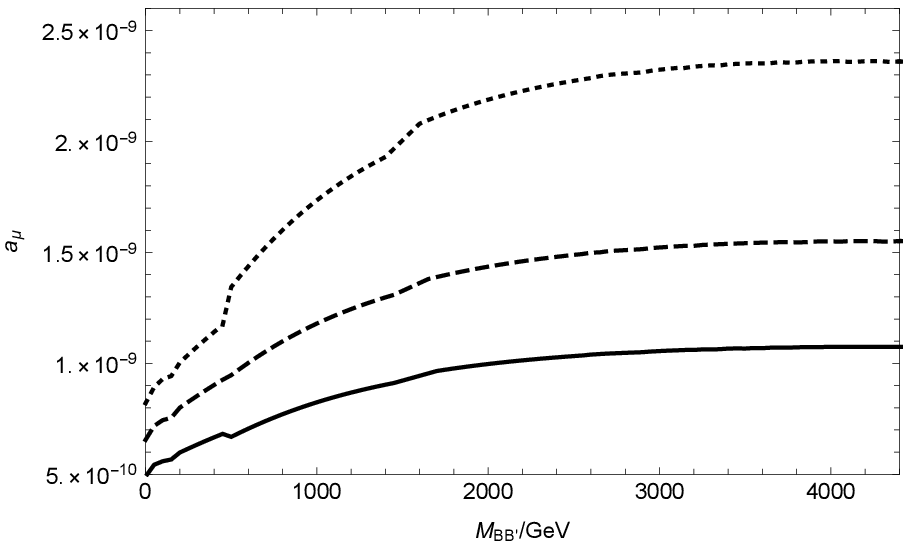}
\end{minipage}%
\caption{$a_\mu$ versus $M_{BB^\prime}$. The solid (dashed, dotted) line corresponds to the results with $\tan{\beta}=20~(30, ~40)$.} \label{MBBP}
\end{center}
\end{figure}

To find the feature different from MSSM, we study the parameter $M_{BB^\prime}$ effects to muon MDM.
$M_{BB^\prime}$ is the mass of the two U(1) gauginos mixing, and appears
as the non-diagonal element of the neutralino mass matrix. In the Eq.(\ref{MIAXBLR}) obtained by MIA,
one can easily find that this contribution is proportional to $\sqrt{x_{BB^\prime}}=\frac{M_{BB^\prime}}{\Lambda}$.
That is to say, $M_{BB^\prime}$ is an important parameter beyond MSSM, and can give new contribution.  We plot the
results  versus $M_{BB^\prime}$ in the Fig.\ref{MBBP} with $M_{NU} = 0.4 ~{\rm TeV}^2, ~M_L= 0.6 ~{\rm TeV}^2$.
In this figure, the solid (dashed, dotted) line corresponds to the results as $\tan{\beta}=20~(30, ~40)$. The three lines are all
increasing functions, when $M_{BB^\prime}$ turns large from 0 to 4400 GeV. The growth trends become weaker and weaker with the increasing $M_{BB^\prime}$, and they are very gentle as $M_{BB^\prime}>3000~{\rm GeV}$. The reason of this feature comes from two
conflict sides: 1 larger $M_{BB^\prime}$ can improve the new contributions;
 2 larger $M_{BB^\prime}$ leads to heavier neutralino, then suppresses the contributions.
 In the region of $M_{BB^\prime}$ from 2000 GeV to 4400 GeV, the dotted line is around $2.2\times10^{-9}$,
 the dashed line is around $1.5\times10^{-9}$,  the solid line is around $1\times10^{-9}$.

\begin{figure}[t]
\begin{center}
\begin{minipage}[c]{0.48\textwidth}
\includegraphics[width=2.9in]{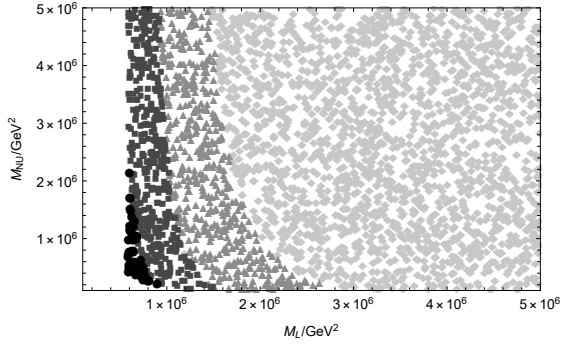}
\end{minipage}%
\caption{$a_\mu$ in the plane of $M_L$ versus $M_{NU}$.} \label{MLMNU}
\end{center}
\end{figure}

To scan the parameter space better, with $\tan\beta = 40$ and $M_{BB^\prime} = 3 ~{\rm TeV}$, we show $a_\mu$ in the
plane of $M_L$ versus $M_{NU}$ in the Fig.\ref{MLMNU}. $M_L$ and $M_{NU}$
affect the masses of scalar leptons and scalar neutrinos. So, they should influence $a_\mu$ to some extent.
The light-gray lozenge \textcolor{light-gray} {$\blacklozenge$} represents the results as $0<a_\mu<10^{-9}$.
The gray triangle \textcolor{mid-gray}{$\blacktriangle$} denotes the results in the region $10^{-9}\leq a_\mu<1.5\times10^{-9}$.
The dark-gray square \textcolor{dark-gray}{$\blacksquare$} denotes the results in the region $1.5\times10^{-9}\leq a_\mu<2\times10^{-9}$.
The filled circle $\bullet$ represents the results as $2\times10^{-9}\leq a_\mu<3\times10^{-9}$.
What needs illustration is that the represented values of \textcolor{light-gray} {$\blacklozenge$}, \textcolor{mid-gray}{$\blacktriangle$}, \textcolor{dark-gray}{$\blacksquare$} and $\bullet$ are also suitable for the following numerical results.
In this figure, \textcolor{light-gray} {$\blacklozenge$} takes up a lot of space.
Similar as the feature of MSSM, heavy scalar lepton and heavy scalar neutrino suppress the SUSY contributions to muon MDM.
It is easy to see that \textcolor{light-gray} {$\blacklozenge$}, \textcolor{mid-gray}{$\blacktriangle$}, \textcolor{dark-gray}{$\blacksquare$} and $\bullet$ are obviously layered. $\bullet$ concentrates in the narrow area
$M_L~(0.75,1){\rm TeV}^2$ and $M_{NU}~(0.2,1.7){\rm TeV}^2$. The blank area as $M_L<0.75~{\rm TeV}^2$
can give large contributions to $a_\mu$, but it is excluded by the scalar lepton mass constraint from LHC.
One can also find that $M_L$ are more sensitive than $M_{NU}$, because $M_L$ affect the masses of both scalar lepton and
scalar neutrino, and $M_{NU}$ just influence scalar neutrino masses.

\begin{figure}[t]
\begin{center}
\begin{minipage}[c]{0.48\textwidth}
\includegraphics[width=2.9in]{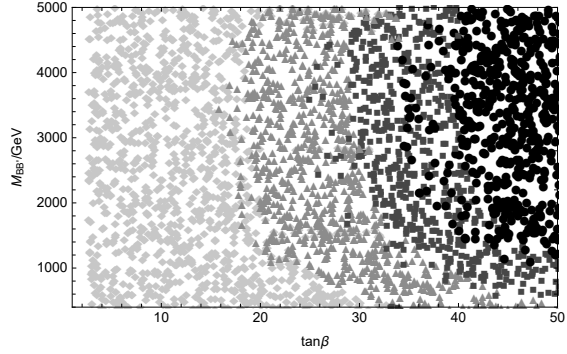}
\end{minipage}%
\caption{$a_\mu$ in the plane of $\tan\beta$ versus $M_{BB^\prime}$.} \label{TBMBBP}
\end{center}
\end{figure}

As $M_L= 0.6 ~{\rm TeV}^2,~ M_{NU} = 0.4 ~{\rm TeV}^2$,
$a_\mu$ in the plane of $\tan\beta$ versus $M_{BB^\prime}$ is shown by
the Fig.\ref{TBMBBP}. When $\tan\beta<20$, the space is filled with \textcolor{light-gray} {$\blacklozenge$}.
In the range $20\leq\tan\beta\leq30$, \textcolor{mid-gray}{$\blacktriangle$}  occupies much space.
$\bullet$ denoting large contribution to $a_\mu$
concentrates in the area $38<\tan\beta<50$ and $1400 ~{\rm GeV}<M_{BB^\prime}<5000 ~{\rm GeV}$.
The results imply that large $\tan\beta$ and large $M_{BB^\prime}$ produce suitable SUSY corrections to compensate the departure.
The bounds between \textcolor{light-gray} {$\blacklozenge$}, \textcolor{mid-gray}{$\blacktriangle$}
and \textcolor{dark-gray}{$\blacksquare$} are obvious.

\subsection{The numerical results with many variables}

In order to analyse the results more extensively, we calculate $a_\mu$ numerically with the scanned parameters:
$1\leq\tan\beta\leq50,~0.2 ~{\rm TeV} \leq M_1 \leq 5 ~{\rm TeV}, ~0.2 ~{\rm TeV} \leq M_{BL} \leq 5
~{\rm TeV},~0\leq M_{BB^\prime} \leq 5 ~{\rm TeV},~0.2\leq g_X\leq 0.6,~0.01\leq g_{YX} \leq 0.5$,
 $0.1 ~{\rm TeV}^2 \leq M_L \leq 10 ~{\rm TeV}^2,~0.1
 ~{\rm TeV}^2 \leq M_E \leq 10 ~{\rm TeV}^2,~0.1 ~{\rm TeV}^2 \leq M_{NU} \leq 10 ~{\rm TeV}^2$,
 $-5~{\rm TeV}\leq T_{nu}\leq 5~{\rm TeV}$. These parameters include sensitive parameters and insensitive parameters.
 So we show the results in several groups of the parameters to find the laws.

\begin{figure}[t]
\begin{center}
\begin{minipage}[c]{1.0\textwidth}
\includegraphics[width=2.9in]{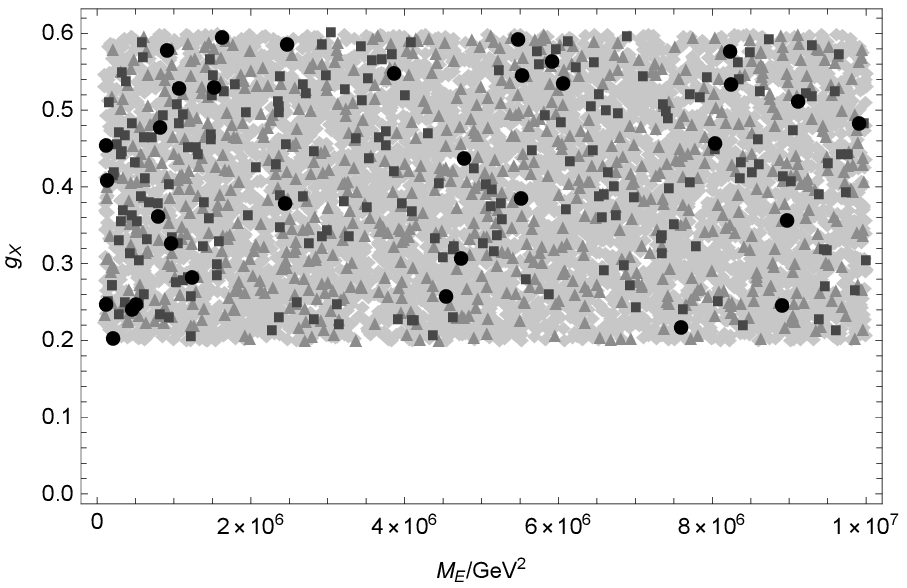}~~~~~~\includegraphics[width=2.9in]{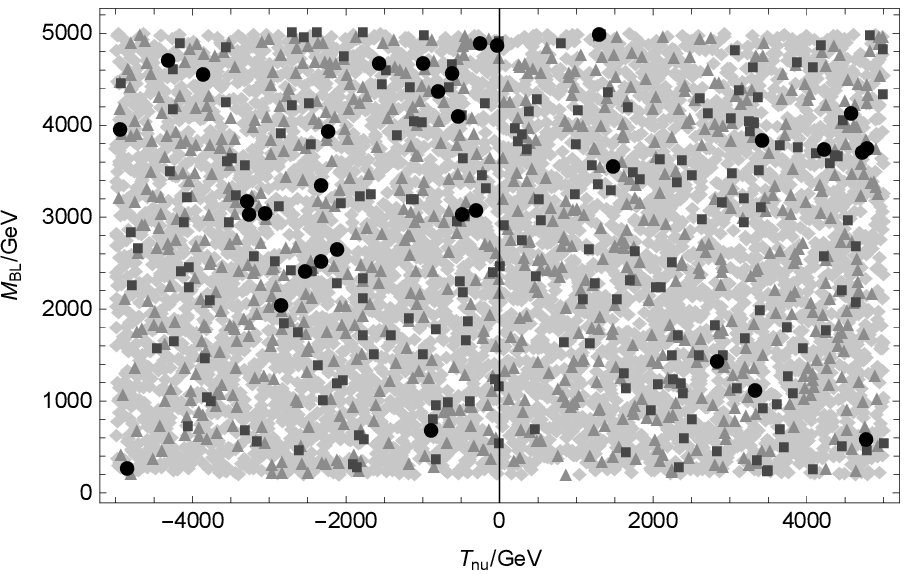}
\end{minipage}%
\caption{The left diagram denotes $a_\mu$ in the plane of $M_E$ versus $g_X$.
The right diagram denotes $a_\mu$ in the plane of $T_{nu}$ versus $M_{BL}$.} \label{Bumingxian}
\end{center}
\end{figure}

 In the Fig.\ref{Bumingxian}, the left diagram shows the results in the plane of $M_E$ versus $g_X$,
where we can not find obvious rule for the results.
\textcolor{light-gray}{$\blacklozenge$}, \textcolor{mid-gray}{$\blacktriangle$}, \textcolor{dark-gray}{$\blacksquare$} and $\bullet$
are distributed in a disorderly way in the plane. The right diagram in the Fig.\ref{Bumingxian} represents the results in the plane of $T_{nu}$ versus $M_{BL}$, that possesses similar character as the left diagram.
These results imply that the effects to $a_\mu$ from $M_E$, $g_X$, $T_{nu}$ and $M_{BL}$ are gentle.

\begin{figure}[t]
\begin{center}
\begin{minipage}[c]{1.0\textwidth}
\includegraphics[width=2.9in]{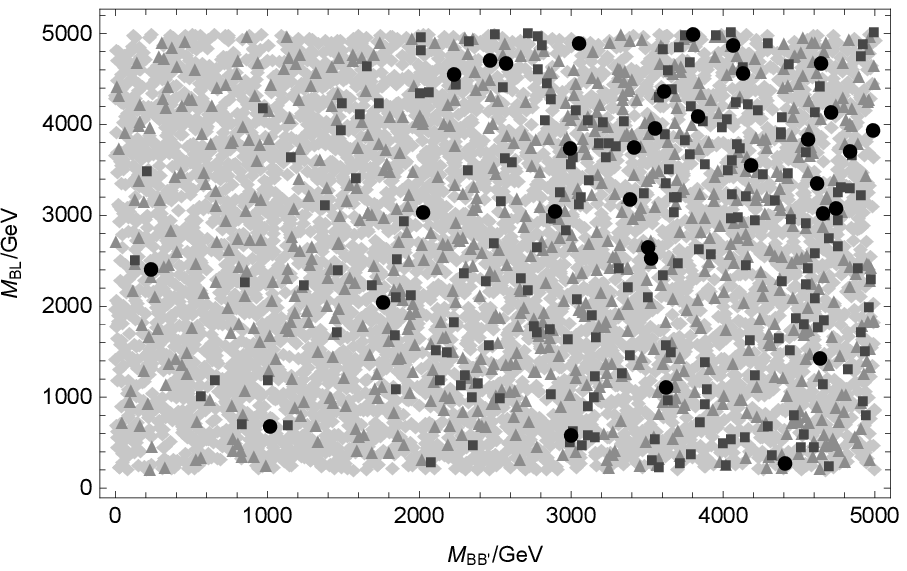}~~~~~~\includegraphics[width=2.9in]{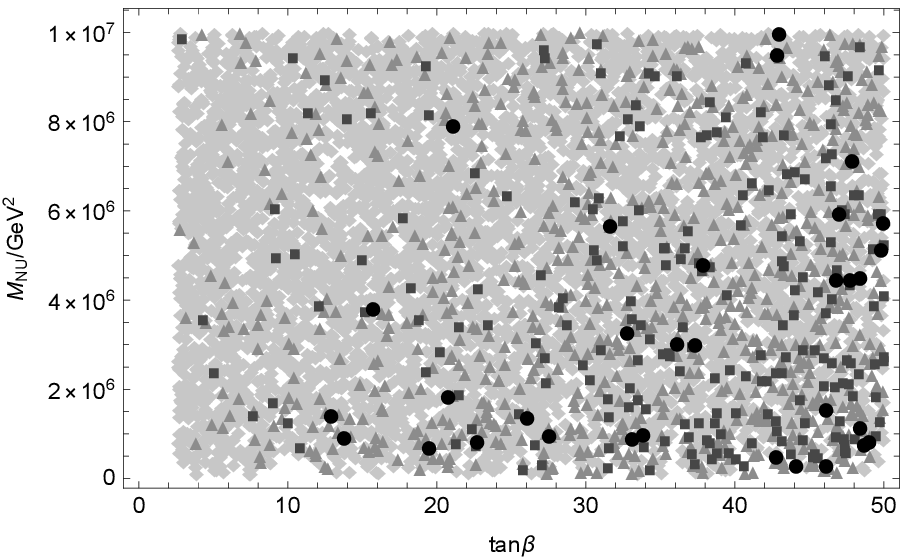}
\end{minipage}%
\caption{The left diagram denotes $a_\mu$ in the plane of $M_{BB^\prime}$ versus $M_{BL}$.
The right diagram denotes $a_\mu$ in the plane of $\tan\beta$ versus $M_{NU}$.} \label{yiban}
\end{center}
\end{figure}

We plot $a_\mu$ in the plane of $M_{BB^\prime}$ versus $M_{BL}$ by the left diagram in the Fig.\ref{yiban},
while the right diagram shows the relation between $a_\mu$, $\tan\beta$ and $M_{NU}$.
The both diagrams in the Fig.\ref{yiban} reflect the common law, though it is not very clear.
In the left diagram, there more $\bullet$ in the top right corner. When $M_{BB^\prime} < 2000 {\rm GeV}$,
the color of the figure is light gray. We can conclude that
$M_{BB^\prime}$ is a sensitive parameter and $M_{BL}$ is a dull parameter.
In the right diagram, larger corrections (darker area) appear at the lower right corner.
It shows that large $\tan\beta$ and small $M_{NU}$ can improve the theoretical corrections.
This rule is consistent with the case of MSSM.
 When $M_{NU}$ turn larger and $\tan\beta$ becomes smaller, the color of the diagram turns lighter.
Large $M_{NU}$ lead to heavy scalar neutrino and suppress the new physics contribution.

\begin{figure}[t]
\begin{center}
\begin{minipage}[c]{1.0\textwidth}
\includegraphics[width=2.9in]{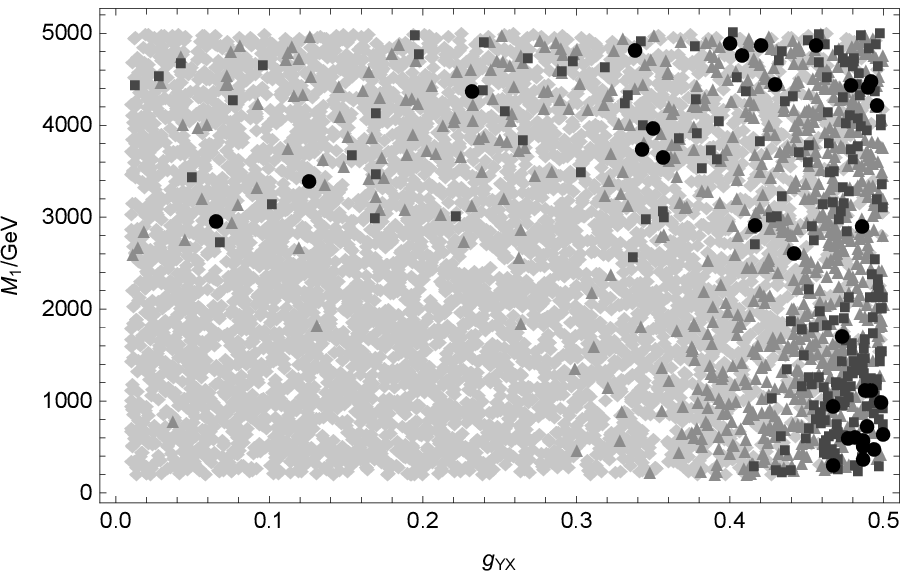},~~~~~~\includegraphics[width=2.9in]{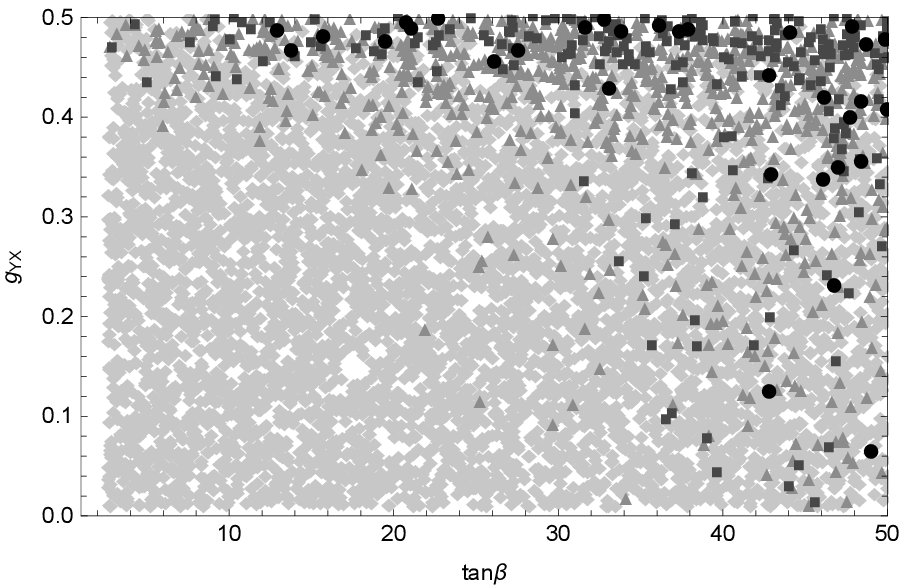}
\end{minipage}%
\caption{The left diagram denotes $a_\mu$ in the plane of $g_{YX}$ versus $M_1$.
The right diagram denotes $a_\mu$ in the plane of $\tan\beta$ versus $g_{YX}$.} \label{hao}
\end{center}
\end{figure}

 In the Fig.\ref{hao}, the left diagram reflects the results in the plane of $M_1$ versus $g_{YX}$.
 $g_{YX}$ is the coupling constant of gauge mixing, and it is the parameter beyond MSSM.
 From the analysis by MIA, $g_{YX}$ is an important parameter.
 In the area $0.4<g_{YX}<0.5$, the color of the diagram is dark.
 When $M_1$ is bigger than 2500 GeV, there are also a certain amount of large results with  $g_{YX}<0.3$.
 \textcolor{light-gray}{$\blacklozenge$} concentrates in the area $g_{YX}<0.3$ and $M_1<2500~{\rm GeV}$.
 $g_{YX}$ and $M_1$ are both sensitive parameters. Furthermore,  $g_{YX}$ is more sensitive than $M_1$.
 The right diagram  reflects the results in the plane of $\tan\beta$ versus $g_{YX}$.
 The both diagrams in the Fig.\ref{hao} are more clear  than the diagrams in the Fig.\ref{yiban} and Fig.\ref{Bumingxian}.
 These results plotted in the right diagram manifest that $\tan\beta$ and $g_{YX}$ are both sensitive parameters.
 There are many \textcolor{dark-gray}{$\blacksquare$} and $\bullet$ in the up side of $g_{YX}$ and right side of $\tan\beta$.
 The top right corner is the most concentrated place for the large results. The bottom left corner
  is denominated by the \textcolor{light-gray}{$\blacklozenge$}. In the whole, large $\tan\beta$ and large $g_{YX}$ can
  obviously improve the corrections to $a_\mu$.

\section{discussion and conclusion}

Extending MSSM with the $U(1)_X$ local gauge group and introducing three Higgs singlets and right-handed neutrinos,
we obtain  $U(1)_X$SSM. In this model, the one-loop diagrams and some important two-loop diagrams are researched  with the effective Lagrangian method. To apparently see the sensitive parameters, the MIA method is used to analyze the one-loop SUSY contributions.
Based on our previous works for the two-loop corrections to muon MDM, the studied two-loop diagrams include: Barr-Zee type, rainbow type and diamond type. It is well known that the one-loop corrections are more important than the two-loop corrections. The works of muon g-2 in MSSM show that
large $\tan\beta$ can improve loop corrections obviously under the constraint for scalar lepton and chargino from LHC.
If all the SUSY particles are very heavy, the loop corrections will be suppressed evidently.
These two characteristics relating with $\tan\beta$ and SUSY particle masses are general in the SUSY corrections to $a_\mu$.

Here, we discuss the  speciality of the $U(1)_X$SSM contributions to $a_\mu$.
$M_{BB^\prime}$ is the mass for the mixing of the $U(1)_Y$ gaugino and $U(1)_X$ gaugino,
and it is the non diagonal element of neutralino mass matrix. Large $M_{BB^\prime}$ can distinctly boost
the one-loop contributions, which is reflected in Eq.(\ref{MIAXBLR}). The gauge mixing coupling constant
$g_{YX}$ is also an important parameter.
From Eqs.(\ref{MIAXLR}), (\ref{MIAHXR}), (\ref{MIAXHL}), (\ref{MIAXBLR}) and the Fig.\ref{hao}, one easily finds that large $g_{YX}$
can improve $a_\mu$ strongly. In the used parameter spaces, the one-loop corrections are dominated.
 The ratio ($a^{2L}_\mu/a^{1L}_\mu$) of just two-loop results to the one-loop results is
 around $10\%$.
 From the numerical results, we find that the corrections from the studied three types of two-loop
diagrams(Barr-Zee type, rainbow type, diamond type) are in the region $10^{-10}\sim10^{-12}$.

In the numerical calculation, we take many parameters as variables including $\tan\beta,~g_X,~g_{YX},~M_1,~M_{BL}, ~M_{BB^\prime}$,
 $ M_L,~M_E,~ M_{NU} $ and $T_{nu}$. The best numerical result of $a_\mu$ is around $2.5\times10^{-9}$,
 which can well compensate the departure between the experiment data and SM prediction.
  Through the analysis of the numerical results, we
 find that $\tan\beta,~M_L,~M_{NU},~M_1,~M_{BB^\prime}$ and $g_{YX}$ are sensitive parameters.
 $a_\mu$ is an increasing function of $\tan\beta,~M_{BB^\prime},~g_{YX}$ and decreasing
 function of $M_L$ and $M_{NU}$. Large $M_L$ and $M_{NU}$ lead to heavy scalar lepton and scalar neutrino,
 then SUSY contributions to $a_\mu$ are suppressed by heavy particles.
  $g_X$, $M_{BL}$, $M_E$ and $T_{nu}$ are insensitive parameters, that give mild influences on the numerical results.  There are a great many two-loop diagrams contributing to $a_\mu$, and some of them can also give considerable corrections.
 In the near future, we shall study other important two-loop diagrams for muon MDM.

\begin{acknowledgments}
This work is supported by National Natural Science Foundation of China (NNSFC)
(No. 12075074), Natural Science Foundation of Hebei Province
(A2020201002), and the youth top-notch talent support program of the Hebei Province.
\end{acknowledgments}


\begin{thebibliography}{99}
\bibitem{1948Phys}J.S. Schwinger, Phys. Rev. {\bf73} (1948) 416.
\bibitem{g2rep2020}T. Aoyama, N. Asmussen, M. Benayoun, et al., Phys. Rep. {\bf 887} (2020) 1.
\bibitem{add}
M. Davier, A. Hoecker, B. Malaescu, et al., Eur. Phys. J. C {\bf77} (2017) 827;
A. Kurz, T. Liu, P. Marquard, et al., Phys. Lett. B {\bf734} (2014) 144;
B. Chakraborty, et al., Phys. Rev. Lett. {\bf 120} (2018) 152001;
S. Borsanyi, et al., Phys. Rev. Lett. {\bf 121} (2018) 022002;
D. Giusti, V. Lubicz, G. Martinelli, et al., Phys. Rev. D {\bf99} (2019) 114502;
E. Shintani, Y. Kuramashi, Phys. Rev. D {\bf100} (2019) 034517;
C.T.H. Davies, et al., Phys. Rev. D {\bf101} (2020) 034512;
A. Gerardin, M. Ce, G.V. Hippel, et al., Phys. Rev. D {\bf100} (2019) 014510;
C. Aubin, T. Blum, C. Tu, et al., Phys. Rev. D {\bf101} (2020) 014503;
D. Giusti, S. Simula, PoS LATTICE {\bf2019 } (2019) 104;
K. Melnikov, A. Vainshtein, Phys. Rev. D {\bf70} (2004) 113006;
P. Masjuan, P.S. Puertas, Phys. Rev. D {\bf95} (2017) 054026;
G. Colangelo, M. Hoferichter, M. Procura, et al., JHEP {\bf04} (2017) 161;
M. Hoferichter, B.L. Hoid, B. Kubis, et al., JHEP {\bf10} (2018) 141;
A. Gerardin, H.B. Meyer, A. Nyffeler, Phys. Rev. D {\bf100} (2019) 034520;
J. Bijnens, N.H. Truedsson, A.R. Sanchez, Phys. Lett. B {\bf798} (2019) 134994;
V. Pauk, M. Vanderhaeghen, Eur. Phys. J. C {\bf 74} (2014) 3008;
I. Danilkin, M. Vanderhaeghen, Phys. Rev. D {\bf 95} (2017) 014019;
F. Jegerlehner, Springer Tracts Mod. Phys. {\bf 274} (2017) 1;
M. Knecht, S. Narison, A. Rabemananjara, et al., Phys. Lett. B {\bf 787} (2018) 111;
P. Roig, P.S. Puertas, Phys. Rev. D {\bf101} (2020) 074019;
G. Colangelo, M. Hoferichter, A. Nyffeler, et al., Phys. Lett. B {\bf 735} (2014) 90.
\bibitem{GWB}G.W. Bennett, et al., Phys. Rev. D {\bf73} (2006) 072003.
\bibitem{AKDN1}A. Keshavarzi, D. Nomura, T. Teubner, Phys. Rev. D {\bf97} (2018) 114025.
\bibitem{GCMH}G. Colangelo, M. Hoferichter,  P. Stoffer, JHEP {\bf 02} (2019) 006.
\bibitem{MHBL}M. Hoferichter, B.L. Hoid, B. Kubis, JHEP {\bf 08} (2019) 137.
\bibitem{MDAH}M. Davier, A. Hoecker, B. Malaescu, et al., Eur. Phys. J. C {\bf80} (2020) 241 [Erratum: Eur. Phys. J. C {\bf80} (2020) 410].
\bibitem{AKDN2}A. Keshavarzi, D. Nomura, T. Teubner, Phys. Rev. D {\bf 101} (2020) 014029.

\bibitem{TBPA}T. Blum, P.A. Boyle, V. Gulpers, et al., Phys. Rev. Lett. {\bf121} (2018) 022003.
\bibitem{TAMH}T. Aoyama, M. Hayakawa, T. Kinoshita, et al., Phys. Rev. Lett. {\bf 109} (2012) 111808.
\bibitem{GCFH}G. Colangelo, F. Hagelstein, M. Hoferichter, et al., JHEP {\bf03} (2020) 101.
\bibitem{GECS}G. Eichmann, C.S. Fischer, R. Williams, Phys. Rev. D {\bf 101} (2020) 054015.
\bibitem{TBNC}T. Blum, N. Christ, M. Hayakawa, et al., Phys. Rev. Lett. {\bf124} (2020) 132002.
\bibitem{TATK}T. Aoyama, T. Kinoshita, M. Nio, Atoms {\bf 7} (2019) 28.
\bibitem{ACWJ}A. Czarnecki, W.J. Marciano, A. Vainshtein, Phys. Rev. D {\bf67} (2003) 073006, [Erratum: Phys. Rev. D {\bf73} (2006) 119901].
\bibitem{CGDS}C. Gnendiger, D. Stockinger, H.S. Kim, Phys. Rev. D {\bf 88} (2013) 053005.


\bibitem{had2}M.T. Hansen, A. Patella,  JHEP {\bf 10} (2020) 029.
\bibitem{muon2}H. Davoudiasl, W.J. Marciano, Phys. Rev. D {\bf98} (2018) 075011.
\bibitem{mdm2} K. Hagiwara, A. Keshavarzi, A.D. Martin, et al., Nucl. Part. Phys. Proc. {\bf287-288} (2017) 33-38.

\bibitem{fnal}Muon g-2 Collaboration, Phys. Rev. D {\bf103} (2021) 072002.

\bibitem{wx1}M. Endo, K. Hamaguchi, S. Iwamoto, et al., JHEP {\bf 07} (2021) 075, arXiv:2104.03217.

\bibitem{wx2} M. Chakraborti, L. Roszkowski, S. Trojanowski, JHEP {\bf 05} (2021) 252, arXiv:2104.04458.

\bibitem{wx3}F. Wang, L. Wu, Y. Xiao, et al., Nucl. Phys. B {\bf 970} (2021) 115486,  arXiv:2104.03262.

\bibitem{wx4}P. Cox, C.C. Han, T.T. Yanagida, Phys. Rev. D {\bf 104} (2021) 075035, arXiv:2104.03290.
\bibitem{wx5}M.V. Beekveld, W. Beenakker, M. Schutten, et al., arXiv:2104.03245.
\bibitem{wx6}M. Chakraborti, S. Heinemeyer, I. Saha, IFT-UAM/CSIC-21-033, arXiv:2104.03287.
\bibitem{wx7}P. Athron, C. Balazs, D.HJ Jacob, et al., JHEP {\bf09} (2021) 080, arXiv:2104.03691.

\bibitem{046}Muon g-2 Collaboration, Phys. Rev. Lett. {\bf126} (2021) 141801.

\bibitem{susy1}S. Heinemeyer, D. St\"{o}ckinger, G. Weiglein, Nucl. Phys. B {\bf690} (2004) 62.
\bibitem{susy2}S. Heinemeyer, D. St\"{o}ckinger, G. Weiglein, Nucl. Phys. B {\bf699} (2004) 103.

\bibitem{caojj}J.J. Cao, J.W. Lian, Y.S. Pan, et al., arXiv: 2104.03284.
\bibitem{shuj}R.Y. Zhou, L.G. Bian, J. Shu, arXiv: 2104.03519.

\bibitem{s1}A. Arhrib, S. Baek, Phys. Rev. D {\bf65} (2002) 075002.
\bibitem{s2}S. Marchetti, S. Mertens, U. Nierste, et al., Phys. Rev. D {\bf79} (2009) 013010.
\bibitem{s3} P.V. Weitershausen, M. Schafer, H.S. Kim, et al., Phys. Rev. D {\bf81} (2010) 093004.
\bibitem{s4}  H.G. Fargnoli, C. Gnendiger, S. Passehr, et al., Phys. Lett. B {\bf726} (2013) 717-724.
\bibitem{s5} H. Fargnoli, C. Gnendiger, S. Passehr, et al., JHEP {\bf02} (2014) 070.
\bibitem{s6} P. Athron, M. Bach, H.G. Fargnoli, et al., Eur. Phys. J. C {\bf76} (2016) 62.

\bibitem{two1}S.M. Zhao, F. Wang, B. Chen, et al., MPLA {\bf28} (2013) 1350173.
\bibitem{two2}S.M. Zhao, T.F. Feng, T. Li, et al.,  MPLA {\bf27} (2012) 1250045.
 \bibitem{other1}G. Degrassi, G. Giudice, Phys. Rev. D {\bf58} (1998) 053007.
 \bibitem{other3}T.F. Feng, L. Sun, X.Y. Yang, Phys. Rev. D {\bf77} (2008) 116008.
 \bibitem{other2}T.F. Feng, X.Y. Yang, Nucl. Phys. B {\bf814} (2009) 101.
\bibitem{one}S.M. Zhao, T.F. Feng, H.B. Zhang, et al., JHEP {\bf11} (2014) 119.

\bibitem{MSSM15fit} C. Strege, G. Bertone, G.J. Besjes,  et al., JHEP {\bf 09} (2014) 081.

\bibitem{MSSM11fit} E. Bagnaschi, K. Sakurai, M. Borsato, et al., Eur. Phys. J. C {\bf 78} (2018) 256.

\bibitem{MSSMgfit} P. Athron, C. Balazs, T. Bringmann, et al.,  Eur. Phys. J. C {\bf77} (2017) 879.

\bibitem{MSSMg2DM} M. Chakraborti, S. Heinemeyer, I. Saha, IFT--UAM/CSIC--21-022, arXiv:2103.13403.

\bibitem{Sarah}F. Staub, SARAH, (2008) arXiv: 0806.0538.
\bibitem{ZSMJHEP}S.M. Zhao, T.F. Feng, M.J. Zhang, et al., JHEP {\bf02} (2020) 130.
\bibitem{slh}L.H. Su, S.M. Zhao, X.X. Dong, et al., Eur. Phys. J. C {\bf81} (2021) 433.
\bibitem{model1}G. Belanger, J.D. Silva, H.M. Tran, Phys. Rev. D {\bf95} (2017) 115017.
\bibitem{model2}V. Barger, P.F. Perez, S. Spinner, Phys. Rev. Lett. {\bf102} (2009) 181802.
\bibitem{model3}P.H. Chankowski, S. Pokorski, J. Wagner, Eur. Phys. J. C {\bf47} (2006) 187.
\bibitem{ffa}X.Y. Yang, T.F. Feng, Phys. Lett. B {\bf675} (2009) 43.
\bibitem{lepton}T.F. Feng, L. Sun, X.Y. Yang, Nucl. Phys. B {\bf800} (2008) 221-252.
\bibitem{our}S.M. Zhao, X.X. Dong, L.H. Su, et al., Eur. Phys. J. C {\bf80} (2020) 823.
\bibitem{dabeta1} T. Moroi, Phys. Rev. D {\bf 53} (1996) 6565-6575; Phys. Rev. D {\bf56} (1997) 4424 (Erratum), hep-ph/9512396.
\bibitem{dabeta2} D. Stockinger, J. Phys. G {\bf34} (2007) R45-R92, hep-ph/0609168.

\bibitem{MIAhao} E. Arganda, M.J. Herrero, R. Morales, et al., JHEP 03 (2016) 055, hep-ph/1510.04685.

\bibitem{zhkrp} R. Grigjanis, P.J.O. Donnell, M. Sutherland, et al., Phys. Rept. {\bf 228} (1993) 93-173.
\bibitem{fengtf04} T.F. Feng, Phys. Rev. D {\bf70} (2004) 096012.

 \bibitem{npb93tl}A.I. Davydychev, J.B. Tazsk, Nucl. Phys. B {\bf 397} (1993) 123.
\bibitem{2020pdg} Particle Data Group, Prog. Theor. Exp. Phys. {\bf 2020} (2020) 083C01.
\bibitem{Zp5d1}The ATLAS Collaboration, Phys. Lett. B {\bf796} (2019) 68, arXiv:1903.06248.
\bibitem{UPbmzgx}G. Cacciapaglia, C. Cs\'{a}ki, G. Marandella, et al., Phys. Rev. D {\bf 74} (2006) 033011;
M. Carena, A. Daleo, B.A. Dobrescu, et al., Phys. Rev. D {\bf 70} (2004) 093009.

\bibitem{TBnew}L. Basso,  Adv. High Energy Phys. {\bf 2015} (2015) 980687.



\end{thebibliography}
\end{document}